\documentclass[11pt]{article}
\pdfoutput=1
\usepackage{jcapmod}
\usepackage{booktabs}
\usepackage[english]{babel}
\usepackage{amsmath,amssymb,amsbsy,amstext,amsthm, simplewick,nicefrac}
\usepackage{hyperref}
\usepackage{graphicx}
\usepackage{amsfonts}
\usepackage{amssymb}
\usepackage{amsmath}
\usepackage{autobreak}
\usepackage{color}
\allowdisplaybreaks[4]
\usepackage{exscale,relsize}
\usepackage[makeroom]{cancel}
\usepackage{soul}
\usepackage{float}
\usepackage[utf8]{inputenc}
\RequirePackage{color}
\usepackage{breqn}

\usepackage{colortbl}
\definecolor{green2}{cmyk}{0, 1, 0.5, 0}
\definecolor{lightgreen}{cmyk}{0.2, 0, 0.2, 0.2}
\definecolor{lightgray}{cmyk}{0.1,0.2,0,0.1}
\definecolor{lightgray2}{cmyk}{0.4,0.4,0,0.8}
\definecolor{black}{cmyk}{1.0,1.0,1.0,1.0}

\allowdisplaybreaks[1]

\usepackage{colortbl}

\makeatletter
\newlength{\apb@width}
\newcommand{\autoparbox}[2][c]{\settowidth{\apb@width}{#2}\parbox[#1]{\apb@width}{#2}}

\makeatother

\numberwithin{equation}{section}

\def\be{\begin{equation}}
\def\ee{\end{equation}}

\def\bea{\begin{eqnarray}}
\def\eea{\end{eqnarray}}
\def\bse{\begin{dmath}}
\def\ese{\end{dmath}}

\def\d{{\rm d}}

\def\d{{\rm d}}

\def\l{{\ell}}
\def\a{{\rm a}}
\def\b{{\rm b}}
\def\H{{\cal H}}

\def\g{{\tt g}}
\def\0{{\boldsymbol 0}}

\def\v{{\boldsymbol{v}}}
\def\x{{\boldsymbol{x}}}

\def\u{{\boldsymbol{u}}}

\DeclareSymbolFont{extraup}{U}{zavm}{m}{n}
\DeclareMathSymbol{\varheart}{\mathalpha}{extraup}{86}
\DeclareMathSymbol{\vardiamond}{\mathalpha}{extraup}{87}

\definecolor{olive}{rgb}{0.5, 0.5, 0.0}
\definecolor{gray}{rgb}{0.5, 0.5, 0.5}

\definecolor{green}{rgb}{0.0, 0.5, 0.0}
\definecolor{auburn}{rgb}{0.43, 0.21, 0.1}
\definecolor{darkviolet}{rgb}{0.58, 0.0, 0.83}
\definecolor{jazzberryjam}{rgb}{0.65, 0.04, 0.37}
\definecolor{mediumviolet-red}{rgb}{0.78, 0.08, 0.52}

\def\a{\alpha}        
\def\b{\beta}       
\def\g{\gamma}    
\def\d{\delta}

\def\k{\kappa}
\def\l{{\lambda}} \def\L{\Lambda}
\def\m{\mu} \def\n{\nu}
\def\o{\omega}
 
\def\r{\rho}
\def\s{\sigma}  \def\S{\Sigma}
\def\t{\tau}

\def\z{\zeta}

\def\pa{\partial}

\DeclareRobustCommand{\SkipTocEntry}[4]{}

\begin{document}

\begin{titlepage}
	
	\setcounter{page}{1} \baselineskip=15.5pt \thispagestyle{empty}
	
	\bigskip\
	
	\vspace{1cm}
	\begin{center}
		
		{\fontsize{20}{28}\selectfont  \sffamily \bfseries 
		 Unimodular vs Nilpotent Superfield Approach to
		 Pure dS Supergravity 
		 }
		
	\end{center}

	\vspace{0.2cm}
	
	\begin{center}
		{\fontsize{13}{30}\selectfont \bf Sukruti Bansal$^{a,}$\footnote{sukruti.b@chula.ac.th}, Silvia Nagy$^{b,c,}$\footnote{s.nagy@qmul.ac.uk}, Antonio Padilla$^{b,}$\footnote{antonio.padilla@nottingham.ac.uk}, Ivonne Zavala$^{d,}$\footnote{e.i.zavalacarrasco@swansea.ac.uk}
			}
	\end{center}
\setcounter{footnote}{0} 
	
	\begin{center}
		
		\vskip 8pt
		\textsl{$^a$ 
		Department of Physics, Faculty of Science, Chulalongkorn University, Phayathai Rd.,
Bangkok 10330, Thailand}\\
		\textsl{$^b$ 
		School of Physics and Astronomy, University of Nottingham, Nottingham NG7 2RD, UK}\\
		\textsl{$^c$ Centre for Research in String Theory, School of
Physics and Astronomy, \\
Queen Mary University of London, 327 Mile End
Road, London E1 4NS, UK}\\
		\textsl{$^d$ Department of Physics, Swansea University, Swansea, SA2 8PP, UK}
		\vskip 7pt
		
	\end{center}

	\vspace{1.2cm}
	\hrule \vspace{0.3cm}
	\noindent {\sffamily \bfseries Abstract} \\[0.1cm]
	
	Recent progress in understanding de Sitter spacetime in supergravity and string theory has led to the development of a four dimensional supergravity with spontaneously broken supersymmetry allowing for de Sitter vacua, also called de Sitter supergravity. 
One approach makes use of constrained (nilpotent) superfields, while an alternative one couples supergravity to a   locally supersymmetric generalization of the Volkov-Akulov goldstino action. These two approaches have been shown to give rise to the same 4D action. 
A novel approach to de Sitter vacua in supergravity involves the  generalisation of unimodular gravity to supergravity using a super-St\"uckelberg mechanism. 
In this paper, we make a connection between   this new approach and the previous two which are in the context of nilpotent superfields and the goldstino brane. We show that upon appropriate field redefinitions, the 4D actions match up to the cubic order in the fields. This points at the possible existence of a more general framework to obtain de Sitter spacetimes from high-energy theories.    
	\vskip 10pt
	\hrule
	
	\vspace{0.6cm}
\end{titlepage}

\tableofcontents

\section{Introduction}\label{Sec1}

In recent years there has been significant work on the role of the Volkov-Akulov (VA) goldstino  in  the spontaneous breaking of local supersymmetry and the generation of a positive contribution to the cosmological constant in supergravity \cite{Dudas:2015eha,Bergshoeff:2015tra,Hasegawa:2015bza,Ferrara:2015gta,Kuzenko:2015yxa,Antoniadis:2015ala}.  In \cite{Bergshoeff:2015tra} the construction of  dS supergravity is based on a constrained (nilpotent) superfield description of the VA action as described  for example in \cite{Rocek:1978nb,Lindstrom:1979kq,Casalbuoni:1988xh,Komargodski:2009rz,Kuzenko:2011tj}. 
In the  nilpotent superfield description, the scalar partners of the goldstino   are not elementary but composed from goldstino bilinears. An alternative approach to constructing  four-dimensional supergravity with spontaneously broken supersymmetry allowing for de Sitter vacua was considered in \cite{Bandos:2015xnf}. The authors considered a locally supersymmetric generalisation of the VA goldstino action describing the dynamics of a space-filling non-BPS 3-brane in 4D ${\cal N}=1$ superspace coupled to the superspace action of minimal ${\cal N}= 1$ 4D supergravity. They then show that to  quadratic order in the goldstino field the action they get coincides with the  supergravity constructions using nilpotent superfields. 

In \cite{Nagy:2019ywi} three of us took another, very different, approach to dS solutions in supergravity.  Specifically, we considered a non-trivial generalisation of unimodular gravity to minimal ${\cal N} = 1$ 4D supergravity\footnote{For other supersymmetric extensions of the unimodular theory see \cite{Nishino:2001gd,Baulieu:2020obv,Anero:2019ldx,Anero:2020tnl}.}, which allowed for de Sitter solutions.
In particular, we formulated  a superspace version of the St\"uckelberg mechanism, which then restored diffeomorphism and local supersymmetry invariance. The cosmological constant and gravitino mass then arise naturally as the vacuum expectation values (vevs) of the components of a Lagrange multiplier superfield after imposing the super-unimodularity condition. At first sight our work \cite{Nagy:2019ywi} is qualitatively different from the constructions using the goldstino brane or equivalently the nilpotent superfield constructions \cite{Bandos:2015xnf,Bergshoeff:2015tra}. However, it is natural to ask if there is a relationship between our approach and that of \cite{Bandos:2015xnf,Bergshoeff:2015tra}. This is a step towards understanding whether there is a unique action describing dS in pure supergravity. In this paper we set about answering this question. Although these constructions are all non-perturbative, a direct comparison at that level is very difficult. Therefore, we will do this perturbatively up to the third order in the relevant fields.  

In the next section, we briefly highlight the main results of the construction in \cite{Bandos:2015xnf}, which we use to connect the results  from unimodular supergravity constructions to pure dS supergravity constructions. We then present a brief review of our unimodular supergravity construction  \cite{Nagy:2019ywi} in \autoref{review_unimod_sugra}. In \autoref{unimod_sugra_cubic} we perturb the unimodular supergravity action up to the cubic order. In \autoref{Goldstino Dynamics} we describe the goldstino dynamics up to the cubic order in unimodular supergravity, which is compared with the goldstino brane action in \cite{Bandos:2015xnf}. We summarise and discuss our results in \autoref{Disc}. Our conventions are summarised in the appendices.

\section{The Goldstino Brane Action in Supergravity}\label{Sec2}

In this section we briefly highlight  the main results of the construction of dS supergravity in \cite{Bandos:2015xnf}. The approach there was to couple a local generalisation of the VA action  to minimal ${\cal N}=1$ supergravity formulated in superspace \cite{Wess:1978bu,Wess:1992cp,Buchbinder:1998qv}. 
The starting action in superspace was given by\footnote{There is an overall minus sign difference with \cite{Bandos:2015xnf}  also used in \cite{Bandos:2016xyu}.}
\be\label{bandos1}
S_B= - \frac{3}{2\kappa^2} \int {d^8 z \,{\rm Ber}\, E} - \frac{2\, m^{\tiny (B)}}{\kappa^2} \left(\int{d^6\zeta_L {\cal E} + {\rm h.c.}}\right) - f^2\!\!\int{d^4\xi} \,{\rm det} {\mathbb E}(z(\xi))\,,
\ee
where $\kappa^2=8\pi G_N$ and $G_N$ is the gravitational constant, 
$z^M=(x^\mu,\theta^\alpha,\bar\theta^{\dot\beta})$ are the coordinates of superspace whose curved geometry is described by the supervielbein
$E^A=dz^M E_M^A(z)$, which contains the fields of the minimal supergravity multiplet \cite{Wess:1992cp} and  Ber $E$ is the Berezinian of $E_M^A$.
The superscript $^{(B)}$ denotes the fields used in \cite{Bandos:2015xnf}. This is done to differentiate them from the fields used in our work \cite{Nagy:2019ywi}, and the dictionary between the two is given in \autoref{App1}. In \eqref{bandos1}, the first term is the action of standard old minimal supergravity, while the second term gives  the anti-de-Sitter cosmological constant term and ${\cal E}$ is the volume measure of the chiral subspace, $\zeta_L^{\cal M} =(x_L^\mu,\Theta^\alpha)$. The third term in \eqref{bandos1} is the generalisation of  the VA  goldstino action, giving  the full non-linear contribution of the 3-brane. Its coupling to supergravity is given via its embedding in the bulk superspace as
\be
\xi^i \to z^M(\xi)= (x^\mu(\xi),\theta^\alpha(\xi), \bar\theta^{\dot\beta}(\xi))\,,
\ee
where $\xi^i $ are the 3-brane worldvolume coordinates with $i=0,1,2,3$. Finally, det${\mathbb E}(z(\xi))$ denotes the determinant of the pullback of the vector supervielbein, 
$E^a(z)=dz^M E^a_M$,  that is, ${\mathbb E}_i^a(z(\xi))\equiv \partial_iE^a(z(\xi))$, while the parameter $f^2$ denotes the 3-brane tension, giving a positive contribution to the cosmological constant.
 
Action \eqref{bandos1} can be rewritten in chiral subspace using the identity 
 $\int{d^6\zeta_L{\cal E}} = 2\int{d^8z\frac{{\rm Ber} E}{{\cal R}}}$, where $\cal R$ is the 
 superspace curvature superfield (see appendix \ref{App2} for its components), as follows: 
 \be\label{bandosch}
 S_B= -\frac{1}{\kappa^2} \int{d^6\zeta_L {\cal E} \left(6 {\cal R}   + 2\, m^{\tiny (B)} + {\rm h.c.}\right)} - f^2\int{d^4\xi} \,{\rm det} {\mathbb E}(z(\xi))\,,
 \ee
 where the explicit expression for $\cal E$ is shown later in eqs.~\eqref{volume_density}. The component form of action \eqref{bandos1} can be derived by integrating the first two terms over the Grassmann odd coordinates and fixing the static-gauge $x^\mu(\xi)=\delta_i^\mu\xi^i$ on the 3-brane. This gives the action \cite{Bandos:2015xnf}

\be\label{Stot}
S_B = S_{SG} + S_{VA}\,,
\ee
with 
\allowdisplaybreaks
\begin{align}
S_{SG} &= -\frac{1}{2\kappa^2} \int d^4x \, e\, \big[  R^{(B)} - 4\,e^{-1} \varepsilon^{\mu\nu\eta\lambda}\left({\cal D}_\nu\psi^{(B)}_\eta\sigma_\lambda\bar\psi_\mu^{(B)}   + \psi_\mu^{(B)}\sigma_\nu{\cal D}_\rho \bar\psi^{(B)}_\lambda  \right)\nonumber\\
&\qquad\qquad\qquad\quad\ \ - 4m^{(B)}(\bar\psi^{(B)a} \sigma_{ab}\bar \psi^{(B)b}+\psi^{(B)a} \sigma_{ab}\psi^{(B)b})+\tfrac{3}{32}\, b_a^{\tiny (B)}b^{{\tiny (B)}\,a} \nonumber\\
&\qquad\qquad\qquad\quad\ \ +\tfrac{3}{8} \left(4m^{(B)}+M^{\tiny (B)}\right)\left(4m^{(B)}+\bar M^{\tiny (B)}\right) -6(m^{(B)})^2\big]\,, \label{Ssg}\\
S_{VA} &=- f^2 \int d^4x \,{\rm det} \,{\mathbb E}(x,\theta(x),\bar\theta(x)) \,. \label{SVA}
\end{align}

The coupling of the VA goldstino to the supergravity fields is encoded in $S_{VA}$, where $\theta^\alpha(x)$ is the VA goldstino. $S_{SG}$ gives standard AdS supergravity, where $e=\text{det}\,e^a_\mu$, $e_\mu^a(x)$ is the spacetime vielbein, $\psi^{(B)}_\mu(x)$ is the gravitino, $R^{(B)}$ is the scalar curvature, ${\cal D}_\m$ denotes the covariant derivative and $b_a^{(B)}$ and $M^{(B)}$ are the  minimal supergravity  auxiliary fields. 

Computation of $S_{VA}$ up to the quadratic order in the fermionic fields gives
\begin{align}
S_{VA}= -f^2\int d^4 x \, e \,[&1+2i\left(\theta\sigma^a\bar\psi_a^{(B)} - \psi_a^{(B)}\sigma^a\bar\theta\right)+ i \left(\theta\sigma^a{\cal D}_a\bar\theta-{\cal D}_a\theta\sigma^a\bar\theta\right) -\tfrac18 b_a^{\tiny (B)}\theta\sigma^a\bar\theta 
\nonumber\\
&+\tfrac12\left(\theta\theta \bar M^{(B)} + \bar\theta\bar\theta M^{(B)}\right) +\dots]
\end{align}
Substituting this into  \eqref{Stot} and varying with respect to the auxiliary fields $b_a^{(B)}$ and $M^{(B)}$, one can find the solutions to their field equations. 
Substituting these solutions back into action \eqref{Stot} one finally gets, up to the second order in the fermions \cite{Bandos:2015xnf}: 
\begin{align}\label{bandos2}
   S_B=-\int d^4x\,e\,&\big\{ \frac{1}{2\kappa^2}\,[R^{(B)}-4\,e^{-1}\,\varepsilon^{\mu\nu\eta\lambda}({\cal D}_\nu\psi_\eta^{(B)}\sigma_\lambda\bar\psi_\mu^{(B)}+ \psi_\mu^{(B)}\sigma_\nu{\cal D}_\rho\bar\psi_\lambda^{(B)})\nonumber\\
&-4m^{(B)}(\bar\psi^{(B)a}\sigma_{ab}\bar\psi^{(B)b} + \psi^{(B)a}\sigma_{ab}\psi^{(B)b})]+\big(f^2- 3\frac{(m^{(B)})^2}{\kappa^2}\big) \nonumber \\
    &+f^2[2i(\theta\sigma^a\bar\psi^{(B)}_a - \psi^{(B)}_a\sigma^a\bar\theta) + 
      i \left(\theta\sigma^a{\cal D}_a\bar\theta - 
      {\cal D}_a\theta\sigma^a\bar\theta\right)-2m^{(B)} \left(\theta^2+\bar\theta^2\right)
      \!] +\dots\big\}
\end{align}
In order to compare our results in  \cite{Nagy:2019ywi} with those of \cite{Bandos:2015xnf,Bergshoeff:2015tra}, we  perturb this action up to the 3rd order in all the field fluctuations, which gives:
\begin{align}\label{Bandos_pert}
    S
    =\frac{1}{2\kappa^2}\int d^4x\,\big[&\{-eR^{(B)}+4\,\varepsilon^{\m\n\eta\l}(\psi^{(B)}_\m\sigma_\n\mathcal{D}_\eta\bar\psi^{(B)}_\l+h.c.)\}^{(3)}-\{e(2\k^2f^2-6(m^{(B)})^2)\}^{(3)} \nonumber \\
    &+\{-4m^{(B)}\bar\psi^{(B)a}\sigma_{ab}\bar\psi^{(B)b} -f^2(2i\theta\sigma^a\bar\psi^{(B)}_a +i\theta\sigma^a\mathbf{D}_a\bar\theta 
       - 2m^{(B)}\theta^2) \nonumber\\
    &+2\,hm^{(B)}\psi^{(B)}_a\sigma^{ab}\psi^{(B)}_b+4m^{(B)}\psi^{(B)}_ah_c{}^{[a}\sigma^{b]c}\psi^{(B)}_b +2\kappa^2f^2(-ih\theta\sigma^a\bar\psi^{(B)}_a \nonumber\\
    &+i\theta h^a{}_b\sigma^b\bar\psi^{(B)}_a -\tfrac{i}{2}h\theta\sigma^a \mathbf{D}_a\bar\theta+\tfrac{i}{2}\theta h^a{}_b\sigma^b \mathbf{D}_a\bar\theta +i\theta\sigma^a\bar{\omega}^{(1)}_a\bar\theta + hm^{(B)}\theta^2) + h.c.\}\big]\,,
\end{align}
where ${\bf D}_\m$ is the covariant derivative on the background and 
we use $\{...\}^{(3)}$  to denote a perturbation up to the cubic order. We use this notation throughout the paper for the perturbation of the standard part of supergravity. Using the dictionary between the notations in \cite{Nagy:2019ywi} and \cite{Bandos:2015xnf} given in \autoref{App1}, we get the following action:
\begin{align}\label{bandos3}
    S=\frac{1}{2\kappa^2}\int d^4x\,\big[\big\{&\sqrt{-g}\,R-\varepsilon^{\mu\nu\rho\lambda}\left(\psi_\m\sigma_\n \mathcal{D}_\r\bar\psi_\l+h.c.\right)\big\}^{(3)}-\big\{2\sqrt{-g}\,\left(\mathbf{\Lambda}_2-\frac{m^2}{3}\right)\big\}^{(3)} \nonumber \\
    &+\big\{\tfrac{2}{3} m\psi_\mu\sigma^{\mu\nu}\psi_\nu+2i\mathbf{\Lambda_2}\breve{\mathcal{G}}\sigma^\mu\bar{\psi}_\mu-2i\mathbf{\Lambda_2}\breve{\mathcal{G}}\sigma^\mu\mathbf{D}_\mu\bar{\breve{\mathcal{G}}}+\tfrac{4}{3} m\mathbf{\Lambda_2}\breve{\mathcal{G}}^2 \nonumber\\
    &+\,\tfrac13 hm\psi_\mu\sigma^{\mu\nu}\psi_\nu+\tfrac23m\psi_\mu h_\rho{}^{[\mu}{} \sigma^{\nu]\rho}\psi_\nu+\mathbf{\L}_2
    ( ih\breve{\mathcal{G}}\sigma^\mu\bar\psi_\mu -  i\breve{\mathcal{G}}h^\mu{}_\nu\sigma^\nu\bar \psi_\mu \nonumber \\
    & -ih\breve{\mathcal{G}}\sigma^\mu{\bf D}_\mu \bar{\breve{\mathcal{G}}}+i\breve{\mathcal{G}}h^\mu{}_\nu\sigma^\nu{\bf D}_\mu\bar{\breve{\mathcal{G}}}+2i\breve{\mathcal{G}}\sigma^\mu\bar\omega_\mu^{(1)}\bar{\breve{\mathcal{G}}}+\tfrac23hm\breve{\mathcal{G}}^2)+h.c.\big\}\big]\,,
\end{align}
where we have used $ \sqrt{-g}=e$. This is the action with which we will compare our final result in \autoref{Goldstino Dynamics}, where the relations of the fields and parameters in \eqref{Bandos_pert} and \eqref{bandos3} will become clear. For example, ${\bf \Lambda}_2$, the vev of the F-term of the Lagrange multiplier in \cite{Nagy:2019ywi}, will be identified with $\kappa^2 f^2$ from \cite{Bandos:2015xnf} and  $\breve{\mathcal{G}}$ is the goldstino.

\section{Review of Unimodular Supergravity}\label{review_unimod_sugra}

\subsection{Unimodular Gravity and the St\"uckelberg Procedure}\label{Review of the Stuckelberg procedure}
The  St\"uckelberg procedure \cite{Stueckelberg:1900zz} is a method for reinstating a broken symmetry into the action at the cost of introducing new fields on which the symmetry is realised nonlinearly. We will first illustrate it here in the setting of unimodular gravity which constitutes the starting point for our construction in \cite{Nagy:2019ywi}. Unimodular gravity \cite{ng,Buch1,Buch2,HT, Kuchar,Ellis,Fiol,meUMG} is a restricted version of Einstein-Hilbert gravity, in which the determinant of the metric is fixed to a constant. The most significant difference between unimodular gravity and GR is the way the cosmological constant enters the theory: as an integration constant rather than appearing directly in the action. The action for unimodular gravity in the absence of matter can be written as
\be 
\label{classic_unim}
\!\!\!S=\frac{1}{16 \pi G_N} \!\!\int d^4x\left[\sqrt{-g}R-2\Lambda(x)\left(\sqrt{-g}-\epsilon_0 \right) \right] \ ,
\ee
with $\epsilon_0$ being a constant (traditionally set to unity, hence the name ``unimodular"). Here $\Lambda(x)$ is a Lagrange multipler field, imposing a local constraint on the determinant of the metric. This version of the Lagrangian is not invariant under the full diffeomorphism group (Diff) but only under a subgroup of transformations called transverse diffeomorphisms (TDiff), whose parameter\footnote{Not to be confused with the brane coordinate used in the previous section!} $\xi_\mu^T$ satisfies $\nabla^\mu\xi_\mu^T=0$. Since the Lagrange multiplier,  $\Lambda(x)$, transforms like a scalar, it is obvious that Diffs are broken by the last term $\int d^4x\Lambda\epsilon_0$.

We now perform the St\"uckelberg trick by transforming the Lagrange multiplier under the \textit{full} group of diffeomorphisms. The transformation rule, written perturbatively in the transformation parameter $\xi^\m$, is:
\be
\Lambda\to\Lambda'=\Lambda-\xi^\m\partial_\m\Lambda+\tfrac{1}{2}\xi^\n\partial_\n\left(\xi^\mu\partial_\mu\Lambda\right)+... 
\ee
As explained above, the action will not be invariant under this, i.e. $S[\Lambda']$ will be a function of $\xi^\mu$. We now promote the transformation parameter to a field: $\xi^\mu\to \phi^\mu$. The Lagrangian can then be constructed order by order in the St\"uckelberg field $\phi^\mu$:  
\be
\label{Lag_uni_sec_ord}
\begin{aligned}
S_\phi=\tfrac{1}{16\pi G_N}\int d^4x\left[\sqrt{-g}R-2\Lambda\sqrt{-g}+ 2\Lambda\epsilon_0\left[1+\partial_\mu \phi^\mu+\tfrac{1}{2}\phi^\n\partial_\n\partial_\mu \phi^\mu+\tfrac{1}{2}\left(\partial_\mu \phi^\mu\right)\left(\partial_\n \phi^\n\right)+... \right]\right] \,,
\end{aligned}
\ee
and it will be invariant, up to the relevant order, when $\phi^\mu$ transforms as\footnote{Note that we are using $\xi$ again to denote the transformation parameter, after performing the St\"uckelberg trick.}:
\be
\delta \phi^\mu=-\xi^\mu-\tfrac{1}{2}\xi^\n\partial_\n \phi^\mu+\tfrac{1}{2}\phi^\n\partial_\n \xi^\mu+...  
\ee
 We note that the St\"uckelberg trick above was performed via an \textit{active} transformation - i.e. we are transforming the fields and keeping the coordinates fixed. One could alternatively do it via a \textit{passive} transformation. Non-pertubatively, this involves sending the coordinates
\be
x^\mu\to \hat  x^\mu(x) \xrightarrow[]{\text{St\"uckelberg}}s^\mu(x) 
\ee 
and the  St\"uckelberged action will be
\be 
\label{lag_uni_nonpert}
S_{s}=\tfrac{1}{16\pi G_N}\int d^4x\left[\sqrt{-g}{\cal R}-2\Lambda\left(\sqrt{-g}-\text{Det}\left(\frac{\partial s}{\partial x} \right)\epsilon_0 \right) \right]\,,
\ee
Of course, upon writing $s^\mu=x^\mu+\phi^\mu$ we find that actions \eqref{lag_uni_nonpert} and \eqref{Lag_uni_sec_ord} are identical order by order in $\phi^\mu$.

\subsection{Unimodular Supergravity and the Super-St\"uckelberg Procedure}
We are working in the conventions of \cite{Wess:1992cp} (see also \autoref{App2}). We find it convenient to work in chiral superspace, where the action of pure $\mathcal{N}=1$ supergravity is given by \be
\label{pure_sugra_superfield_action}
S=-\tfrac{6}{8\pi G_N}\int d^4x\, d^2\Theta\,\mathcal{E} \mathcal{R}+h.c. 
\ee
The components of ${\cal R}$ are given in \autoref{App2} and 
\be
\label{volume_density}
\begin{aligned}
\mathcal{E}&=\mathcal{F}_0\ +\ \sqrt{2}\,\Theta\mathcal{F}_1\ +\ \Theta\Theta \mathcal{F}_2,\qquad\text{with} \\
\mathcal{F}_0&=\tfrac{1}{2}e \,,\\
\mathcal{F}_1&=\tfrac{i\sqrt{2}}{4}e\sigma^\mu\bar{\psi}_\mu \,,\\
\mathcal{F}_2&=-\tfrac{1}{2}eM^*-\tfrac{1}{8}e\bar{\psi}_\mu\left(\bar{\sigma}^\mu\sigma^\nu-\bar{\sigma}^\nu\sigma^\mu \right)\bar{\psi}_\nu\,,
\end{aligned}
\ee
where  $e=\text{det}\,e^a_\mu$, with $e^a_\mu$ the vielbein, $\psi_\mu$ the gravitino and $M$ the scalar auxiliary field in the old minimal supergravity model. We note that $\mathcal{E}$ is the natural supersymmetrisation of the measure $\sqrt{-g}$. It is a chiral density superfield, characterised by the transformation law
\be 
\delta\mathcal{E}=-\partial_N\left[\left(-1\right)^N\eta^N\mathcal{E}\right]\,,\quad
\text{where}\quad
\left(-1\right)^N= 
\left\{\begin{matrix}
1, & N=\mu\\ 
-1, &N=\alpha 
\end{matrix}\right.
\ee
and
\be
\label{etas_def}
\begin{aligned}
\eta^\mu(\epsilon)&=\Theta^\beta y^\mu_{1\beta}(\epsilon) + \Theta^2 y^\mu_2(\epsilon) \,,\\
\eta^\alpha(\epsilon)&=\epsilon^\alpha+\Theta^\beta\Gamma^\alpha_{1\beta}(\epsilon)+\Theta^2\Gamma^\alpha_2(\epsilon)\,,
\end{aligned}
\ee
where $\epsilon$ is the parameter of local SUSY transformations. For conciseness, we introduced the following notation
\be
\label{y_gamma_notation}
\begin{aligned}
y^\mu_{1\alpha}(\epsilon)=&2i\left(\sigma^\mu\bar{\epsilon}\right)_\alpha\,,\\ 
y^\mu_2(\epsilon)=&\bar{\psi}_\nu\bar{\sigma}^\mu\sigma^\nu\bar{\epsilon}\,,\\
\Gamma^\alpha_{1\beta}\left(\epsilon\right)=&-i\left(\sigma^\mu\bar{\epsilon}\right)_\beta\psi_\mu^\alpha\,,\\
\Gamma^\alpha_2(\epsilon)=&-i\omega_\mu^{\alpha\beta}\left(\sigma^\mu\bar{\epsilon}\right)_\beta
+\tfrac{1}{3}M^*\epsilon^\alpha 
-\tfrac{1}{2}\psi_\nu^\alpha\left(\bar{\psi}_\mu\bar{\sigma}^\nu\sigma^\mu\bar{\epsilon}\right) 
+\tfrac{1}{6}b_\mu\left(\varepsilon\sigma^\mu\bar{\epsilon}\right)^\alpha \,,
\end{aligned} 
\ee
where $\omega_\mu^{\alpha\beta}$ is the spin connection and $b_\mu$ is the vector auxiliary field in the old minimal model.  

We define the unimodular supergravity action to be:
\be 
\label{unimodular_sugra_original_action}
S=-\tfrac{6}{8\pi G_N}\int d^4x\, d^2\Theta\left[\mathcal{E} \,\mathcal{R} +\tfrac{1}{6}\Lambda\left(\mathcal{E}-\mathcal{E}_0\right)  \right]        +h.c.
\ee
where 
\be
\Lambda=\Lambda_0+\sqrt{2}\,\Theta \Lambda_1+\Lambda_2\Theta^2 
\ee
is now a Lagrange multiplier chiral superfield, and we defined
\be
\label{curlye0}
\mathcal{E}_0=\epsilon_0 -\tfrac{1}{2} m\Theta^2 \,,
\ee
where $\epsilon_0$ and $m$ are real constants and  the spinor component of $\mathcal{E}_0$ was set to zero for simplicity. Varying over $\Lambda$, we get 
\be
\label{contraint_uni_susy}
\mathcal{E}=\mathcal{E}_0\,.
\ee 
Action \eqref{unimodular_sugra_original_action} is invariant under a restricted set of SUSY and diffeomorphism transformations such that they preserve condition \eqref{contraint_uni_susy}. We remark here that the constraints imposed in eq.~\eqref{contraint_uni_susy} amount to more than a gauge fixing; in this sense, unimodular supergravity is more than a naive supersymmetrisation of the unimodular gravity model in eq.~\eqref{classic_unim}. As a consequence, even though the space of solutions of unimodular gravity matches that of standard Einstein gravity with a cosmological constant, it will not be the case for our model. As we will see in \autoref{Cosmological constant and boundary conditions}, we are no longer restricted to AdS and flat space backgrounds as one is in standard pure supergravity.
Finally, we impose the following boundary conditions on the components of our Lagrange multiplier superfield:
\be
\label{boundary_cond_L_mult}
\Lambda_0\big\vert_{\infty}=K_0\,,\qquad
\Lambda_1\big\vert_{\infty} =0\,, \qquad
\Lambda_2\big\vert_{\infty}=K_2\,,
\ee
where $K_0$ and $K_2$ are some constants. We will see in \autoref{Cosmological constant and boundary conditions} how $K_0$ and $K_2$ become intimately linked with the cosmological constant in our model. 

We are now ready to perform the St\"uckelberg trick in a manner analogous to \autoref{Review of the Stuckelberg procedure}. In components, the symmetry breaking terms following from action \eqref{unimodular_sugra_original_action} are:
\be
 \tfrac{1}{16\pi G_N}\int d^4 x\left[ 2\Lambda_2 \epsilon_0  -m\Lambda_0  +h.c. \right] \ \subset \  S\,.
\ee
We want to perform the Super-St\"uckelberg trick up to the $2^{\text{nd}}$ order in the SUSY and diffeomorphism transformation parameters, so we need to know the $2^{\text{nd}}$ order transformation of the chiral superfield in curved space. This was derived in \cite{Nagy:2019ywi} (working up to the $2^{\text{nd}}$ order in the fermions):
\allowdisplaybreaks
\begin{align}\label{active_transf_lambda}
\delta \Lambda_0=& -\xi^\mu\partial_\mu\Lambda_0-\sqrt{2}\epsilon\Lambda_{1} 
+\tfrac{1}{2}\xi^\n\partial_\n\left(\xi^\mu\partial_\mu\Lambda_0\right)
+\tfrac{1}{2}\epsilon y^\mu_1(\epsilon)\partial_\mu\Lambda_0 
+\tfrac{\sqrt{2}}{2}\xi^\mu\partial_\mu\left(\epsilon\Lambda_1 \right) \nonumber\\
&+\tfrac{\sqrt{2}}{2}\xi^\mu\epsilon\partial_\mu\Lambda_1
+\Lambda_2\epsilon^2  \,, \nonumber\\
\delta \Lambda_1=&-\xi^\mu\partial_\mu\Lambda_1-\tfrac{\sqrt{2}}{2}y_1^\mu(\epsilon)\partial_\mu\Lambda_0
-\sqrt{2}\Lambda_2\epsilon 
+\tfrac{\sqrt{2}}{4}\xi^\n\partial_\n\left(y_1^\mu(\epsilon)\partial_\mu\Lambda_0\right)+\tfrac{\sqrt{2}}{4}y_1^\n(\epsilon)\partial_\n\left( \xi^\mu\partial_\mu\Lambda_0\right) \nonumber\\
&+\tfrac{1}{2}\xi^\n\partial_\n\left(\xi^\mu\partial_\mu \Lambda_1\right)+
\tfrac{\sqrt{2}}{2}\xi^\n\partial_\n\left(\epsilon\Lambda_2\right)
+\tfrac{\sqrt{2}}{2}\epsilon\xi^\mu\partial_\mu\Lambda_2
+\left[\tfrac{i\sqrt{2}}{2}\xi^\n\partial_\n[\sigma^\mu]\bar{\epsilon}-\tfrac{\sqrt{2}}{4}\partial_\n\xi^\mu y^\n_1(\epsilon)\right]\partial_\mu\Lambda_0 \,, \nonumber\\
\delta \Lambda_2=&-\xi^\mu\partial_\mu\Lambda_2+\tfrac{\sqrt{2}}{2}y^{\mu}_1(\epsilon)\partial_\mu\Lambda_{1}-y^\mu_2(\epsilon)\partial_\mu\Lambda_0-Tr(\Gamma_{1}(\epsilon))\Lambda_2
-\sqrt{2}\Gamma_2(\epsilon)  \Lambda_{1} \nonumber\\
&+\tfrac{1}{2}\xi^\n\partial_\n \big[\xi^\mu\partial_\mu\Lambda_2-\tfrac{\sqrt{2}}{2}y^{\mu}_1(\epsilon)\partial_\mu\Lambda_{1}+y^\mu_2(\epsilon)\partial_\mu\Lambda_0
+Tr(\Gamma_{1}(\epsilon))\Lambda_2+\sqrt{2}\Gamma_2(\epsilon) \Lambda_{1}\big] \nonumber\\
&-\tfrac{1}{4}y^{\n}_1(\epsilon)\partial_\n\big[\sqrt{2}\xi^\mu\partial_\mu\Lambda_{1}
+y^\mu_{1}(\epsilon)\partial_\mu\Lambda_0+2\epsilon\Lambda_2\big]
+\tfrac{1}{2}y_2^\n(\epsilon)\partial_\n[\phi^\mu\partial_\mu\Lambda_0]
+\tfrac{1}{2}Tr(\Gamma_{1}(\epsilon))[\xi^\mu\partial_\mu\Lambda_2] \nonumber\\
&+\tfrac{ \sqrt{2}}{2}\Gamma_2(\epsilon)\xi^\mu\partial_\mu\Lambda_{1}
+\tfrac{\sqrt{2}}{4}\big[  -  \xi^\eta \partial_\eta \big[ 2i\sigma^\mu\big]\bar{\epsilon} + \partial_\n \xi^\mu y^{\n}_{1}(\epsilon)\big]
\partial_\mu\Lambda_{1a}
-\tfrac{1}{2}\big[ -\tfrac{1}{2}y^{\nu}_1(\epsilon) \mathcal{D}_\nu y^\mu_{1}(\epsilon) \nonumber\\
& -\xi^\n\partial_\n \big[ \bar{\psi}_\nu\bar{\sigma}^\mu\sigma^\nu\big]\bar{\epsilon}+\partial_\n \xi^\mu y^\n_2(\epsilon) - \tfrac{4i}{3}M^*\epsilon\sigma^\mu\bar{\epsilon}
 -ib_\nu\bar{\epsilon}\bar{\sigma}^\mu\sigma^\nu\bar{\epsilon}+i b_\nu\bar{\epsilon}\bar{\sigma}^\nu\sigma^\mu\bar{\epsilon}
\big]\partial_\mu\Lambda_0 \nonumber\\
&+\tfrac{1}{2} \big[\partial_\mu\epsilon y_{1}^\mu(\epsilon)+
    \xi^\n\partial_\n\big[ i\psi_\mu\sigma^\mu\big]\bar{\epsilon}
+\tfrac{2}{3}M^*\epsilon^2 +\tfrac{4}{3}M\bar{\epsilon}^2 
\big] \Lambda_2 
-\tfrac{\sqrt{2}}{2}\big[  \xi^\n\partial_\n \big[ i\omega_\mu\sigma^\mu  \big]\bar{\epsilon}  - \tfrac{1}{3}\xi^\n\partial_\n M^* \epsilon \nonumber\\
& -\tfrac{1}{6}\xi^\n\partial_\n\big[ b_\mu\varepsilon\sigma^\mu\big]\bar{\epsilon}\big]\Lambda_{1} \,.
\end{align}
Then, sending $\xi^\mu\ \to\ \phi^\mu$ and $\epsilon\ \to\ \zeta$ we obtain the following action\footnote{For simplicity we have set $\epsilon_0=\tfrac{1}{2}$.}, which is given up to the second order in the St\"uckelberg fields by:
\begin{align}\label{action_non_pert}
\!\!\!S=\tfrac{1}{16\pi G_N}\int d^4x\big[&\sqrt{-g}\, [R-\tfrac{2}{3}M^*M+\tfrac{2}{3}b^\mu b_\mu
+\varepsilon^{\mu\nu\eta\l}(\bar{\psi}_\mu\bar{\sigma}_\nu\mathcal{D}_\eta\psi_\l
-\psi_\mu\sigma_\nu\mathcal{D}_\eta\bar{\psi}_\l ) ] \nonumber\\
&+\tfrac{1}{2}\sqrt{-g}\,[-2\Lambda_2+\sqrt{2}i\Lambda_1\sigma^\mu\bar{\psi}_\mu
 +2\Lambda_0\left(\bar{\psi}_\mu\bar{\sigma}^{\mu\nu}\bar{\psi}_\nu+M^* \right)\ +h.c.] \nonumber\\
&+\big\{\big[\Lambda_0 \big(\partial_\mu y^\mu_2(\zeta) -m -m\partial_\mu \phi^\mu\big)
-\Lambda_1\big(\tfrac{\sqrt{2}}{2}\partial_\mu y^\mu_{1}(\zeta)
+\sqrt{2}\Gamma_{2}(\zeta)-\sqrt{2}m\zeta\big) \nonumber\\
&+\Lambda_2[1+\partial_\mu \phi^\mu -Tr(\Gamma_{1}(\zeta))]  \big]
+\Lambda_2[\tfrac{1}{2}\partial_\mu\left(\phi^\mu\partial_\nu \phi^\nu - \phi^\mu Tr(\Gamma_{1}(\zeta))\right) \nonumber\\
&-\tfrac{1}{2}\partial_\mu \phi^\mu  Tr(\Gamma_{1}(\zeta))
+\tfrac{1}{2}\partial_\mu\left(\zeta y^\mu_{1}(\zeta)\right)
+\tfrac{1}{2}\phi^\n\partial_\n( i\psi_\mu\sigma^\mu)\bar{\zeta}
+\tfrac{1}{3}M^*\zeta^2+\tfrac{2}{3}M\bar{\zeta}^2
   \nonumber\\
&-m\zeta^2]  -\tfrac{\sqrt{2}}{2}\Lambda_1\big[\tfrac{1}{2}\partial_\mu\big(\phi^\mu\partial_\nu y^\nu_{1}(\zeta)+y^\mu_{1}(\zeta)\partial_\nu \phi^\nu
+2\phi^\mu\Gamma_{2}(\zeta) \big)
+\Gamma_{2}(\zeta)\partial_\mu \phi^\mu \nonumber\\
&-\tfrac{1}{2}\partial_\mu [\phi^\n \partial_\n (2i\sigma^\mu)\bar{\zeta}
-\partial_\n \phi^\mu y^{\n}_{1}(\zeta) ]
 - \phi^\n\partial_\n [ -i\varepsilon\omega_\mu(\sigma^\mu  ]\bar{\zeta}) 
-\tfrac{1}{3}\phi^\n\partial_\n M^*\zeta \nonumber \\
&-\tfrac{1}{6}\phi^\n\partial_\n( b_\mu\varepsilon\sigma^\mu)\bar{\zeta}
 -m\zeta\partial_\mu \phi^\mu -m\partial_\mu\left(\zeta \phi^\mu\right)]+\tfrac{1}{2}\Lambda_0\partial_\mu[\phi^\mu\partial_\nu y^\nu_2(\zeta)+y^\mu_2(\zeta)\partial_\nu \phi^\nu \nonumber\\
&-\tfrac{1}{2}y^{\mu}_1(\zeta)\partial_\nu y^\nu_{1}(\zeta)
-2(\mathcal{D}_\nu\bar{\zeta})\bar{\sigma}^\mu\sigma^\nu\bar{\zeta}-2(\varepsilon\sigma^\mu\bar\zeta) \omega_\nu\sigma^\nu\bar{\zeta}
-\phi^\eta\partial_\eta ( \bar{\psi}_\nu\bar{\sigma}^\mu\sigma^\nu)\bar{\zeta} \nonumber\\
&+\partial_\n \phi^\mu y^\n_2(\zeta)
 -\tfrac{4i}{3}M^*\zeta\sigma^\mu\bar{\zeta}
 -ib_\nu\bar{\zeta}\bar{\sigma}^\mu\sigma^\nu\bar{\zeta}+i b_\nu\bar{\zeta}\bar{\sigma}^\nu\sigma^\mu\bar{\zeta}
-m\phi^\mu\partial_\n \phi^\n \nonumber\\
&  +m\zeta y^\mu_{1}(\zeta) ]+h.c. \big\} \,+... \ \big]\,,
\end{align}
where the ``..." stand for terms at higher orders in the St\"uckelberg fields $\zeta$ and $\phi^\mu$. The action is now invariant under diffeomorphisms and SUSY transformations, provided the St\"uckelberg fields transform analogously to superspace coordinates (see \cite{Nagy:2019ywi} for details).  We remark that the St\"uckelberg trick can alternatively be applied via passive transformations, in which case the action can be written in the form
\be 
\label{full_action_after_stueck}
\!\!\!S=-\tfrac{6}{8\pi G_N}\int d^4x d^2\Theta\left[\mathcal{E} \mathcal{R} +
\tfrac{1}{6}\Lambda\!\left(\!\mathcal{E}-
\text{Ber}\!\left(\frac{\partial \Phi}{\partial X}\right) \mathcal{E}_0\left(\Phi\right)\!\right)  \!\right]        +h.c.\ ,
\ee
where $\text{Ber}\!\left(\frac{\partial \Phi}{\partial X}\right)$ is the Berezinian for the transformation of the superspace coordinates $X^A$  as $X^A \to {X'}^A \xrightarrow[]{\text{St\"uckelberg}}\Phi^A$ (see \cite{Nagy:2019ywi} for details). Of course, upon expanding action \eqref{full_action_after_stueck}, it will coincide with action \eqref{action_non_pert} order-by-order in the St\"uckelberg fields.

Finally, we note that, upon application of the St\"uckelberg procedure, the boundary conditions \eqref{boundary_cond_L_mult} are modified (to linear order in the St\"uckelberg fields) as:
\begin{align}\label{stuecked_bound_cond_non_pert}
&\big[\Lambda_0-\phi^\mu\partial_\mu\Lambda_0-\sqrt{2}\zeta\Lambda_1\big]\big\vert_{\infty}=K_0\,, \nonumber \\
&\big[\Lambda_1-\phi^\mu\partial_\mu\Lambda_1-\sqrt{2}\zeta\Lambda_2+\tfrac{\sqrt{2}}{2}y_1^\mu(\zeta)\partial_\mu\Lambda_0\big]\big\vert_{\infty}=0\,, \nonumber \\
&\big[\Lambda_2-\phi^\mu\partial_\mu\Lambda_2-\tfrac{\sqrt{2}}{2}y^\mu_1(\zeta)\partial_\mu \Lambda_1-\sqrt{2}\Gamma_2(\zeta)\Lambda_1 
-Tr(\Gamma_1(\zeta))\Lambda_2
-y_2^\mu(\zeta)\partial_\mu\Lambda_0\big]\big\vert_{\infty}=K_2\,,
\end{align}
and are now invariant under diffeomorphisms and SUSY transformations.

\subsection{Cosmological Constant and Boundary Conditions}\label{Cosmological constant and boundary conditions}
We are now interested in the backgrounds allowed by our model, in particular the range of values taken by the cosmological constant. Upon computing the equations of motion following from action \eqref{action_non_pert}, we find that they admit the solution
\allowdisplaybreaks
\begin{align}\label{background_sol}
\langle g_{\mu\nu}\rangle&=\bar{g}_{\mu\nu},\quad\text{with}\quad \sqrt{-\bar{g}}=1\,, \nonumber\\
\langle M\rangle&=m\,, \nonumber\\
\langle\Lambda_0\rangle&=\tfrac{2}{3}m=K_0\,, \nonumber\\
\langle\Lambda_2\rangle&=\mathbf{\Lambda}_2=K_2,\quad\text{with}\quad \text{Im}(\Lambda_2)=0\,,
\end{align} 
with all other fields vanishing. The cosmological constant is
\be \label{cc}
c.c.= \mathbf{\Lambda}_2-\tfrac{1}{3}m^2\ =\ K_2-\tfrac{3}{4}K_0^2\,.
\ee
Thus, our model allows for a cosmological constant of either sign, similar to the results in the constrained superfield literature \cite{Dudas:2015eha,Bergshoeff:2015tra,Hasegawa:2015bza,Ferrara:2015gta,Antoniadis:2015ala,DallAgata:2015pdd,Bandos:2015xnf,Farakos:2016hly,Cribiori:2016qif,Bandos:2016xyu}. However, we emphasise that in our approach the cosmological constant appears as the combination of the vevs of the Lagrange multiplier superfield components $\Lambda_0$ and $\Lambda_2$, which are nothing but the boundary conditions imposed in eqs.  \eqref{stuecked_bound_cond_non_pert} (note that for our solution the St\"uckelberg fields vanish on the background).

\section{Unimodular Supergravity at Cubic Order}\label{unimod_sugra_cubic}

In this section, we connect the goldstino brane/nilpotent superfield approach to dS vacua, and unimodular supergravity. To this end we perturb the unimodular supersymmetric action shown in eq.~\eqref{action_non_pert} around the background solution \eqref{background_sol}, up to  $3^{\text{rd}}$ order in field fluctuations. 

The  fluctuations of the fields around the background solution \eqref{background_sol} are:

\be \label{fluct}
\begin{aligned}
g_{\mu\nu}&=\overline{g}_{\mu\nu}+h_{\mu\nu}\,,\\
\o_\m{}^{\a\b}&=\boldsymbol\o_\m{}^{\a\b}+\o^{(1)}_\m{}^{\a\b} \,,\\
\psi_\mu^\a&=0+\psi_\mu^\a\,,\\
b_\mu&=0+b_\mu\,,\\
M&=  m+M\,,\\
\Lambda_0&=\tfrac{2}{3}m+\lambda_0\,,\\
\Lambda_1&=0+\lambda_1\,,\\
\Lambda_2&=\mathbf{\Lambda}_2+\lambda_2, \qquad \text{Im}(\mathbf{\Lambda}_2)=0\,,\\
\phi^\mu&=0+t^\mu\,,\\
\zeta&=0+\chi\,.
\end{aligned}
\ee
The fluctuations of $\sqrt{-g}$ and and of the Pauli matrices, derived from the above  fluctuations, can be found at the end of \autoref{App2}. 

\subsection{Terms Cubic in St\"uckelberg Fields}\label{Terms Cubic in Stuckelberg Fields}
We now use   \eqref{fluct} to perturb the unimodular supergravity action \eqref{action_non_pert} up to the cubic order in fluctuations. 
That action will contain terms of the kind $\mathcal{L}_{(t^3)}$ and $\mathcal{L}_{(t\chi^2)}$, which are cubic in the St\"uckelberg fluctuations. These terms could be computed by extending the non-perturbative action \eqref{action_non_pert} up to the cubic order in the St\"uckelberg fields and then perturbing it. This turns out to be quite cumbersome.

However, as we will see below, this proves unnecessary as it is possible to prove that the cubic fluctuations of these terms vanish perturbatively.

$\mathcal{L}_{(t^3)}$ denotes terms cubic in the fluctuation $t^\mu$, and $\mathcal{L}_{(t\chi^2)}$ denotes terms linear in $t^\mu$ and quadratic in the fermionic St\"uckelberg field $\chi$, together with their hermitian conjugates. 
$\mathcal{L}_{(t^3)}$ is the cubic fluctuation of the subset of the  St\"uckelberg action coming from the diffeomorphism transformation. This can easily be computed explicitly and shown to reduce to a total derivative in the perturbatve expansion (remembering that the vevs of $\L_0$ and $\L_2$ are constant). Taking $t^\mu$ to vanish on the boundary, we have
\begin{align}
    \int d^4x\,\mathcal{L}_{(t^3)}=0\,.
\end{align}
Now let us look at $\mathcal{L}_{(t\chi^2)}$. It is of the form
\be
\mathcal{L}_{(t\chi^2)}=\mathcal{L}^A(t^\mu,\chi,\chi)+\mathcal{L}^B(t^\mu,\chi,\bar{\chi}) \ +h.c.
\ee
where the most general expressions for 
$\mathcal{L}^A(t^\mu,\chi,\chi)$\footnote{Other possible terms either vanish or reduce to the ones shown by using Pauli matrix identities.} and $\mathcal{L}^B(t^\mu,\chi,\bar{\chi})$ are given by:
\begin{align}\label{tchichiansatz}
    \mathcal{L}^A(t^\mu,\chi,\chi)&=A_1t^\m\mathbf{D}_\m(\chi\chi)+A_2 t^\mu \mathbf{D}_\mu\mathbf{D}_\nu \chi \mathbf{D}^\nu\chi+A_3t^\mu\square\chi\mathbf{D}_\mu\chi+A_4 t^\mu (\square\mathbf{D}_\mu\chi)\chi\,, \nonumber\\
    \mathcal{L}^B(t^\mu,\chi,\bar{\chi})&=B_1\,t^\m\chi\s_\m\bar\chi+B_2 t^\mu\mathbf{D}^\nu\chi\s_\mu\mathbf{D}_\nu\bar{\chi}
+B_3t^\mu\square\chi\s_\mu\bar{\chi} 
+B_4t^\mu\mathbf{D}_\mu\chi \s^\nu\mathbf{D}_\nu\bar{\chi} 
+B_5t^\mu\mathbf{D}_\mu\mathbf{D}_\nu\chi\s^\nu\bar{\chi}\,.
\end{align}
Here $A_i$ with $i=1,...4$ and $B_i$ with $i=1,...5$ are a priori arbitrary constants. However, as can be seen from eq.~\eqref{full_action_after_stueck} the St\"uckelberg fields (appearing inside the Berezinian) are always multiplied by $\L$ in the non-perturbative action. This implies that $A_i$ and $B_i$ must be proportional to either $m$ or ${\bf\L}_2$, which are the vevs of the components of the Lagrange multiplier superfield $\L$. Given that $[m]=1$ and $[{\bf\L}_2]=2$, we get $[A_i]\geq1$ and $[B_i]\geq1$, with $[X]$ denoting the mass dimension of $X$. This forbids higher derivative terms in $\mathcal{L}^A$ and $\mathcal{L}^B$ in \eqref{tchichiansatz}. For example, candidate terms in $\mathcal{L}^B$ consisting of four derivatives will have massless coupling constants. Therefore terms of this type are not allowed in $\mathcal{L}^B$.

$\mathcal{L}_{(t\chi^2)}$ transforms under SUSY transformations as 
\be\label{tchiexp}
\delta \mathcal{L}_{(t\chi^2)}=-\mathcal{L}^A(t^\mu,\epsilon,\chi)-\mathcal{L}^A(t^\mu,\bar\epsilon,\bar\chi)-\mathcal{L}^B(t^\mu,\epsilon,\bar{\chi})-\mathcal{L}^B(t^\mu,\bar{\epsilon},\chi) +\mathcal{O}(4)
\ee
Let the Lagrangian in \eqref{action_non_pert} perturbed up to the cubic order in fluctuations be denoted by $\mathcal{L}.$
Since the full action up to the $3^{\text{rd}}$ order in fluctuations,  $\int d^4x(\mathcal{L}+\mathcal{L}_{(t^3)}+\mathcal{L}_{(t\chi^2)})=\int d^4 x(\mathcal{L}+\mathcal{L}_{(t\chi^2)})\,$, is supersymmetric, $\delta \mathcal{L}_{(t\chi^2)}$ should cancel those terms in $\delta \mathcal{L}$ which have tensor structures as in eq. \eqref{tchiexp}. Denoting those terms by $\delta\mathcal{L}'_{(t\chi^2)}$, we thus require
\begin{align}
    \delta \mathcal{L}_{(t\chi^2)}+\delta\mathcal{L}'_{(t\chi^2)}=0\,.
\end{align}
It can be checked by explicit computation that
\begin{align}
    \delta\mathcal{L}'_{(t\chi^2)}=0\,,
\end{align}
 and therefore,
\begin{align}
    \delta \mathcal{L}_{(t\chi^2)}=0\,.
\end{align}
One can check that $\d\mathcal{L}^A(t^\mu,\chi,\chi)+ \d\mathcal{L}^B(t^\mu,\chi,\bar{\chi})\neq0$ and also $\d\mathcal{L}^A(t^\mu,\chi,\chi)\neq 0\neq \d\mathcal{L}^B(t^\mu,\chi,\bar{\chi})$.
So $\delta \mathcal{L}_{(t\chi^2)}$ can be $0$ only if $\mathcal{L}^A(t^\mu,\chi,\chi)=\mathcal{L}^B(t^\mu,\chi,\bar{\chi})=0$. Hence we see that terms cubic in St\"uckelberg fields are not present in the cubic action.

\subsection{Perturbing the Action up to the Cubic Order}

In \cite{Nagy:2019ywi} we identified the goldstino, denoted by $\mathcal{G}$, at quadratic order with the combination
\be 
\mathcal{G}=\tfrac{1}{2}\left(\chi+\tfrac{\sqrt{2}}{2\mathbf{\L}_2}\lambda_1\right)\,,
\ee 
and the orthogonal mode to $\mathcal{G}$, denoted by $\t$, as
\be 
\tau=\tfrac{1}{2}\left(\chi-\tfrac{\sqrt{2}}{2\mathbf{\L}_2}\lambda_1\right).
\ee 
\noindent
Using these identifications and  perturbing action \eqref{action_non_pert} by the  fluctuations \eqref{fluct} up to the $3^{\text{rd}}$ order, we  obtain the action
\begin{align}\label{cubicaction1}
    S=\tfrac{1}{16\pi G_N}\int d^4x&\big[\big(\sqrt{-g}\,\big[R-\tfrac{2}{3}M^*M+\tfrac{2}{3}b^\mu b_\mu +\varepsilon^{\mu\nu\eta\l}\big(\bar{\psi}_\mu\bar{\sigma}_\nu\mathcal{D}_\eta\psi_\l -\psi_\mu\sigma_\nu\mathcal{D}_\eta\bar{\psi}_\l \big) \big]\big)^{(3)} \nonumber\\
    &+\big\{\!\!-\!\left(2\sqrt{-g}\,\left(\bold{\L}_2-\tfrac23\, m^2\right)\right)^{(3)}+\tfrac{2}{3} m\psi_\mu\sigma^{\mu\nu}\psi_\nu+2i\mathbf{\Lambda_2}\mathcal{G}\sigma^\mu\bar{\psi}_\mu\nonumber\\
    &-2i\mathbf{\Lambda_2}\mathcal{G}\sigma^\mu\mathbf{D}_\mu\bar{\mathcal{G}}+\tfrac{4}{3} m\mathbf{\Lambda_2}\mathcal{G}^2+2i\mathbf{\Lambda_2}\tau\sigma^\mu\mathbf{D}_\mu\bar{\tau}-\tfrac{4}{3} m\mathbf{\Lambda_2}\tau^2+\lambda_2\!\left(\partial_\mu t^\mu -\tfrac{1}{2}h\right)\nonumber\\
    &+\lambda_0[M^*- m (\partial_\mu t^\mu  -\tfrac{1}{2}h)]+\tfrac12 (\tfrac12 h^2 - h_{\m\n}h^{\m\n})(\tfrac13 mM^*+\tfrac12\lambda_0m -\tfrac12 \lambda_2)\nonumber\\
    &+\tfrac{i}{2}{\bf\L}_2(\mathcal{G}-\tau)(h\s^\m-h^\m{}_\eta\s^\eta)\bar{\psi}_\m +(\l_0+\tfrac13 hm)\bar\psi_\m\bar\s^{\m\n}\bar\psi_\n \nonumber\\
    &+\tfrac23\, m\bar \psi_\m h^{[\m}{}_\eta\bar \s^{\n]\eta}\bar\psi_\n+\tfrac12 h\l_0 M^*  +\l_0\,\partial_\mu[\bar\psi_\nu\bar\s^\mu\s^\nu(\bar{\mathcal{G}}+\bar\tau)]\nonumber\\
    &+\tfrac{i}{2}\bold{\L}_2[\psi_\mu-2\,\partial_\mu(\mathcal{G}-\tau)]h^\mu{}_\eta\sigma^\eta(\bar{\mathcal{G}}+\bar\tau)-i\l_2\psi_\m\s^\m(\bar{\mathcal{G}}+\bar\tau)\nonumber\\
    & -2\,{\bf{\L}}_2(\mathcal{G}-\tau)\big[\tfrac{i}{2}\,\boldsymbol \o_\m h^\m{}_\eta\s^\eta(\bar{\mathcal{G}}+\bar\tau))-i\o_\m^{(1)}\s^\m(\bar{\mathcal{G}}+\bar\tau) +\tfrac13 M^* (\mathcal{G}+\tau)\nonumber\\
    &+\tfrac16 b_\m \s^\m(\bar{\mathcal{G}}+\bar\tau)\big] +\tfrac12 (\l_2-\l_0 m)\partial_\m(t^\m\partial_\n t^\n)+\tfrac{i}{2} \bold{\L}_2[ t^\m\partial_\m(\psi_\eta\s^\eta)(\bar{\mathcal{G}}+\bar\tau) \nonumber\\
    &-\partial_\m t^\m \big(\psi\s(\bar{\mathcal{G}}+\bar\tau)\big)\big]+\big(\tfrac{\bold{\L}_2}{6}b_\mu-i\partial_\m\l_2\big)(\mathcal{G}+\tau)\s^\m(\bar{\mathcal{G}}+\bar\tau)\nonumber\\
    & +\tfrac13(\bold{\L}_2M^*-2\l_2 m)(\mathcal{G}+\tau)^2 +\tfrac23(\bold{\L}_2 M+\l_2m)(\bar{\mathcal{G}}+\bar\tau)^2\nonumber\\
    &-i\bold{\L}_2(\mathcal{G}-\tau)\big[\partial_\mu\big(2\,\partial_\nu t^\nu\sigma^\mu(\bar{\mathcal{G}}+\bar\tau)+t^\n\sigma^\mu\partial_\nu(\bar{\mathcal{G}}+\bar\tau)-t^\mu(\boldsymbol\omega_\eta\sigma^\eta(\bar{\mathcal{G}}+\bar\tau))\big)\nonumber\\
    &-\partial_\mu t^\mu\boldsymbol\o_\eta\sigma^\eta(\bar{\mathcal{G}}+\bar\tau)\big]+\tfrac{2}{3}\bold{\L}_2 m(\mathcal{G}-\tau)[2\,\partial_\mu t^\mu(\mathcal{G}+\tau)+t^\mu\partial_\mu(\mathcal{G}+\tau))]\nonumber\\
    &-i\bold{\L}_2(\mathcal{G}-\tau) t^\eta\partial_\eta(\boldsymbol\o_\m\s^\m)(\bar{\mathcal{G}}+\bar\tau)-\l_0\partial_\m[(\bar{\mathcal{G}}+\bar\tau)\bar{\s}^\m\partial_\n\big(\s^\n(\bar{\mathcal{G}}+\bar\tau)\big)\nonumber\\
    & -(\mathcal{D}_\nu(\bar{\mathcal{G}}+\bar\tau))\bar{\sigma}^\mu\sigma^\nu(\bar{\mathcal{G}}+\bar\tau)-(\sigma^\mu(\bar{\mathcal{G}}+\bar\tau)) \boldsymbol\omega_\nu\sigma^\nu(\bar{\mathcal{G}}+\bar\tau)\nonumber\\
    &+\tfrac{i}{3} m(\mathcal{G}+\tau)\s^\m(\bar{\mathcal{G}}+\bar\tau)]+h.c.\big\}\,\big]\,.
\end{align}

This action \eqref{cubicaction1} is invariant under the following diffeomorphic transformations (perturbed up to the $2^{\text{nd}}$ order in fluctuations):
\begin{align}
    \delta h_{\mu\nu}&=-\bar{\nabla}_\mu \xi_\nu - \bar{\nabla}_\mu \xi_\nu -\xi^\eta \bar{\nabla}_\eta h_{\mu\nu}+\xi^\eta \bar{\nabla}_\mu h_{\nu\eta}+ \xi^\eta \bar{\nabla}_\nu h_{\mu\eta}\,, \nonumber\\
    \delta\psi^\mu&=-\,\xi^\n\partial_\n\psi^\mu +\partial_\n \xi^\mu\psi^\n\,, \nonumber \\
    \delta M&= -\,\xi^\mu\partial_\mu M\,, \nonumber \\
    \delta \lambda_0 &= -\,\xi^\mu \partial_\mu \lambda_0 \nonumber \\
    \delta \lambda_2 &=-\,\xi^\mu \partial_\mu \lambda_2 \,,\nonumber \\
     \delta \mathcal{G} &= -\,\xi^\mu \partial_\mu \mathcal{G}\,, \nonumber \\
    \delta t^\mu&=-\,\xi^\mu +h^{\mu\nu}\xi_\nu - \tfrac12 \xi^\n\partial_\n t^\mu +\tfrac12 t^\n\partial_\n\xi^\mu \,,\nonumber \\
    \delta \tau&=-\,\xi^\mu\partial_\mu\tau\,.
\end{align}
Furthermore, action \eqref{cubicaction1} is also invariant under the following supersymmetry transformations (perturbed up to the $2^{\text{nd}}$ order in fluctuations):
\begin{align}\label{SUSY_up_to_2nd}
\delta h_{\mu\nu}&= 0 + 2i(\psi_{(\mu} \sigma_{\nu)} \bar{\epsilon}-\epsilon \sigma_{(\mu} \bar{\psi}_{\nu)} ) \,, \nonumber\\
\delta \psi_\mu&=-2\mathcal{\mathbf{D}}_\mu\epsilon + \tfrac{i}{3} m\left(\varepsilon\sigma_\mu\bar{\epsilon}\right) -2\epsilon\omega^{(1)}_\mu+\tfrac{i}{3}M\left(\varepsilon\sigma_\mu\bar{\epsilon} \right)
+\tfrac{i}{6}m h_\mu^{\ \n}\left(\varepsilon  \sigma_\n  \bar{\epsilon}\right) +ib_\mu\epsilon+\tfrac{i}{3}b^\n\epsilon \sigma_\n\bar{\sigma}_\mu\,, \nonumber \\
\delta M&= 0 -\epsilon\sigma^\alpha\bar{\sigma}^\beta \psi_{\alpha\beta} +im\epsilon\sigma^\alpha \bar{\psi}_\alpha \,, \nonumber \\
\delta b_{\alpha\dot{\alpha}}
&= 0 + \epsilon^\delta(\tfrac{3}{4}\bar{\psi}_{\alpha\phantom{\gamma}\delta\dot{\gamma}\dot{\alpha}}^{\phantom{\alpha}\dot{\gamma}}+\tfrac{1}{4}\epsilon_{\delta\alpha}\bar{\psi}^{\gamma\dot{\gamma}}_{\phantom{\alpha\beta}\gamma\dot{\alpha}\dot{\gamma}}-\tfrac{i}{2}m\psi_{\alpha\dot{\alpha}\delta}
)
-\bar{\epsilon}(\tfrac{3}{4}\psi^\gamma_{\phantom{\alpha}\dot{\delta}\gamma\dot{\alpha}\alpha}
+\tfrac{1}{4}\varepsilon_{\dot{\delta}\dot{\alpha}}\psi_{\alpha\phantom{\alpha\beta}\dot{\gamma}\gamma}^{\phantom{\alpha}\dot{\gamma}\gamma}+\tfrac{i}{2}m\bar{\psi}_{\alpha\dot{\alpha}\dot{\delta}})\,, \nonumber \\
\delta\lambda_0&= 0-\sqrt{2}\epsilon\lambda_1 \,, \nonumber\\
\delta\lambda_2&= 0 +\sqrt{2}i\partial_\mu\lambda_1\sigma^\mu \bar{\epsilon}-i\mathbf{\Lambda_2}\psi_\mu\sigma^\mu \bar{\epsilon}+\sqrt{2}i\lambda_1\boldsymbol\o_\mu\sigma^\mu \bar{\epsilon}
-\tfrac{\sqrt{2}}{3}m\epsilon\lambda_1 \,, \nonumber\\
\delta t^\mu&= 0 + i(\mathcal{G}+\t) \sigma^\mu\bar{\epsilon}-i\epsilon \sigma^\mu(\bar{\mathcal{G}}+\bar{\t}) \,, \nonumber\\
\delta\mathcal{G}&={-\epsilon}+\tfrac14 t^\m\partial_\m\epsilon-\tfrac{1}{2\bf{\L_2}}(i\sigma^\mu  \bar{\epsilon}\partial_\mu\lambda_0+\lambda_2\epsilon)\,, \nonumber \\
\delta \t&=0 +\tfrac14 t^\m\partial_\m\epsilon+\tfrac{1}{2\bf{\L_2}}(i\sigma^\mu  \bar{\epsilon}\partial_\mu\lambda_0+\lambda_2\epsilon)\,.
\end{align} 

\subsection{Boundary Conditions and Field Redefinitions}

We next look at the perturbations of the boundary conditions, \eqref{stuecked_bound_cond_non_pert}. 
Using $K_0=\frac23 m$ and $K_2=\mathbf{\L}_2$ from the background solution \eqref{background_sol}, and the fluctuations \eqref{fluct}, 
up to the $2^{\text{nd}}$ order, we obtain:
\be\label{boundary_cond_2nd}
\begin{aligned}
    &\big[ \lambda_0-t^\mu\partial_\mu\lambda_0+\bold{\L}_2(-\,\mathcal{G}^2+2\,\mathcal{G}\t+3\,\t^2)\big]\vert_\infty=0 \,, \\
   &\big[{\t}+\frac{1}{4}t^\m \partial_\m(\mathcal{G}-3\t)+\tfrac{1}{\mathbf{2\Lambda_2}}\l_2(\mathcal{G}+\t)+\tfrac{i}{2\mathbf{\Lambda_2}}\s^\m (\bar{\mathcal{G}}+\bar\t)\partial_\m\l_0\big]\vert_\infty=0 \,,\\
    &\big[\l_2-\,t^\m \partial_\m \l_2+2i\mathbf{\Lambda_2}\s^\m(\bar{\mathcal{G}}+\bar\t)\partial_\m(\mathcal{G}-\t)-i\mathbf{\Lambda_2}\psi_\m{}^\a(\s^\m(\bar{\mathcal{G}}+\bar\t))_\a  \\
    &+2\,i\mathbf{\Lambda_2}\,\boldsymbol\o_\m{}^\a(\s^\m(\bar{\mathcal{G}}+\bar\t))_\a(\mathcal{G}-\t)+\frac{m\mathbf{\Lambda_2}}{3}(-\mathcal{G}^2+2\,\mathcal{G}\t+3\,\t^2+2\,(\bar{\mathcal{G}}+\bar\t)^2)\big]\vert_\infty=0 \,.
\end{aligned}
\ee
These suggest the following field redefinitions:
\be
\label{rho_redefs}
\begin{aligned}
r_0\equiv\, & \lambda_0-t^\mu\partial_\mu\lambda_0+\bold{\L}_2(-\,\mathcal{G}^2+2\,\mathcal{G}\t+3\,\t^2)\,,\\
\rho_1\equiv\, &\t+\frac{1}{4}t^\m \partial_\m(\mathcal{G}-3\t)+\tfrac{1}{\mathbf{2\Lambda_2}}\l_2(\mathcal{G}+\t)+\tfrac{i}{2\mathbf{\Lambda_2}}\s^\m (\bar{\mathcal{G}}+\bar\t)\partial_\m\l_0 \,,\\
r_2\equiv\, &\l_2-\,t^\m \partial_\m \l_2+2i\mathbf{\Lambda_2}\s^\m(\bar{\mathcal{G}}+\bar\t)\partial_\m(\mathcal{G}-\t)-i\mathbf{\Lambda_2}\psi_\m{}^\a(\s^\m(\bar{\mathcal{G}}+\bar\t))_\a \\
    &+2\,i\mathbf{\Lambda_2}\,\boldsymbol\o_\m{}^\a(\s^\m(\bar{\mathcal{G}}+\bar\t))_\a(\mathcal{G}-\t)+\frac{m\mathbf{\Lambda_2}}{3}(-\mathcal{G}^2+2\,\mathcal{G}\t+3\,\t^2+2\,(\bar{\mathcal{G}}+\bar\t)^2)\,,
\end{aligned}
\ee
so that the boundary conditions can now be expressed more simply as
\be\label{r0_bound_cond}
r_0|_\infty=0,\quad \rho_1|_\infty=0 ,\quad r_2|_\infty=0  \,.
\ee
We need to substitute the old fields $\l_0, \t$ and $\l_2$ by the new fields $r_0, \r_1$ and $r_2$. So we invert relations \eqref{rho_redefs} up to the $2^{\text{nd}}$ order, and get,
\allowdisplaybreaks
\begin{align}\label{quadredef}
\lambda_0\equiv\, & r_0+t^\mu\partial_\mu r_0+\bold{\L}_2(\mathcal{G}^2-2\,\mathcal{G}\rho_1-3\,\rho_1^2) \,,\nonumber\\
\t\equiv\, &\rho_1-\frac{1}{4}t^\m \partial_\m(\mathcal{G}-3\rho_1)-\tfrac{1}{\mathbf{2\Lambda_2}}r_2(\mathcal{G}+\rho_1)-\tfrac{i}{2\mathbf{\Lambda_2}}\s^\m (\bar{\mathcal{G}}+\bar\r_1)\partial_\m r_0\,, \nonumber\\
\l_2\equiv\, & r_2+\,t^\m \partial_\m r_2-2i\mathbf{\Lambda_2}\partial_\m(\mathcal{G}-\rho_1)\s^\m(\bar{\mathcal{G}}+\bar\rho_1)+i\mathbf{\Lambda_2}\psi_\m\s^\m(\bar{\mathcal{G}}+\bar\rho_1) \nonumber\\
&-2\,i\mathbf{\Lambda_2}\,(\mathcal{G}-\rho_1)\boldsymbol\o_\m(\s^\m(\bar{\mathcal{G}}+\bar\rho_1))+\frac{m\mathbf{\Lambda_2}}{3}(\mathcal{G}^2-2\,\mathcal{G}\rho_1-3\,\rho_1^2-2\,(\bar{\mathcal{G}}+\bar\rho_1)^2)\,.
\end{align}
Plugging these expressions for the fields $\l_0, \t$ and $\l_2$ into the cubic action \eqref{cubicaction1}, we get the following action 
\begin{align}\label{cubicaction}
S=\tfrac{1}{2\kappa^2}\int d^4x
&\big\{\big[\big(\sqrt{-g}\,\big[R-\tfrac{2}{3}M^*M+\tfrac{2}{3}b^\mu b_\mu + \varepsilon^{\mu\nu\eta\l}\big(\bar{\psi}_\mu\bar{\sigma}_\nu\mathcal{D}_\eta\psi_\l -\psi_\mu\sigma_\nu\mathcal{D}_\eta\bar{\psi}_\l \big) \big]\big\}^{(3)}\nonumber\\
    & +\big\{-\big(2\sqrt{-g}\,(\bold{\L}_2-\tfrac23\, m^2)\big)^{(3)} +\tfrac{2}{3} m\psi_\mu\sigma^{\mu\nu}\psi_\nu+2i\mathbf{\Lambda_2}\mathcal{G}\sigma^\mu(\bar{\psi}_\mu-\pa_\mu\bar{\mathcal{G}})\nonumber\\
    &+\tfrac{4}{3} m\mathbf{\Lambda_2}(\mathcal{G}^2-\r_1^2) +2i\mathbf{\Lambda_2}\r_1\s^\m\pa_\m\bar\r_1  +(-mr_0+r_2)\left(\partial_\n t^\n -\tfrac{1}{2}h\right)+r_0 M^*\nonumber\\
    &+i\mathbf{\Lambda_2}[(\mathcal{G}-3\r_1)\pa_\n(t^\n\s^\m\pa_\m\bar\r_1)-\pa_\n(t^\n\pa_\m(\mathcal{G}-\rho_1))\s^\m(\bar{\mathcal{G}}+\bar{\rho}_1)\nonumber\\
    &+t^\n\partial_\n(\mathcal{G}-3\r_1)\s^\m\bar{\boldsymbol\o}_\m\bar\r_1 +\tfrac{2}{{\bold \L}_2}r_2(\mathcal{G}+\r_1)\s^\m\bar{\boldsymbol\o}_\m\bar\r_1 -\tfrac{2\,i}{{\bold \L}_2}(\mathcal{G}+\r_1)\bar{\s}^\n\partial_\n r_0\s^\m\bar{\boldsymbol\o}_\m\bar\r_1]\nonumber\\
    &+ir_2(\mathcal{G}+\r_1)\s^\m\pa_\m(\bar{\mathcal{G}}-\bar\r_1)+ir_2\pa_\m(\mathcal{G}+\r_1)\s^\m(\bar{\mathcal{G}}+\bar{\r}_1) \nonumber\\
&+ (\mathcal{G}+\r_1)\bar\s^\m\pa_\m r_0\s^\n\pa_\n(\bar{\mathcal{G}}-\bar{\r}_1)+\tfrac{2}{3} m\mathbf{\Lambda_2}[\pa_\m t^\m(\tfrac12 \mathcal{G}^2+3\r_1^2+\mathcal{G}\r_1)-\pa_\m\mathcal{G}t^\m\r_1] \nonumber\\
&-\tfrac23 mr_2[(\mathcal{G}^2-\r_1^2)-(\bar{\mathcal{G}}+\bar\r_1)^2]-\tfrac13 mi(\mathcal{G}-3\r_1)\s^\m(\bar{\mathcal{G}}+\bar{\r}_1)\pa_\m r_0\nonumber\\
    &-M^*r_0\left(\partial_\n t^\n -\tfrac{1}{2}h\right) -\,\tfrac{2}{3}m\mathbf{\Lambda_2}(\mathcal{G}+\r_1)^2\,\partial_\n t^\n+\tfrac23 mh{\bf\L}_2(\mathcal{G}^2-\r_1^2)\nonumber\\
    &+i\mathbf{\Lambda_2}\partial_\m(\mathcal{G}-\rho_1)(h\s^\m-h^\m{}_\n\s^\n)\big(\s^\m(\bar{\mathcal{G}}+\bar\rho_1)\big) -\tfrac12 ht^\m \partial_\m r_2+ih\mathbf{\Lambda_2}\mathcal{G}\s^\m\bar\psi\nonumber\\
    &-\tfrac{i}{2}\mathbf{\Lambda_2}t^\m\psi_\r\s^\r\pa_\m(\bar{\mathcal{G}}+\bar\r_1)+\tfrac23\bold{\L}_2(\mathcal{G}^2-2\,\mathcal{G}\rho_1-3\,\rho_1^2)M^* -t^\m r_0\pa_\m M^*\nonumber\\
    & -\tfrac12 mt^\m \pa_\m r_0(\partial_\n t^\n - h) +\tfrac12 \big(\tfrac12 h^2 - h_{\m\n}h^{\m\n}\big)\big(\tfrac13 mM^*+\tfrac12r_0m-\tfrac12 r_2\big)\nonumber\\
    & +i{\bf\L}_2\psi_\m h^\m{}_\n\s^\n\bar{\mathcal{G}}+(r_0+ \tfrac13 hm)\bar\psi_\m\bar\s^{\m\n}\bar\psi_\n +\tfrac23\, m\bar \psi_\m h^{[\m}{}_\r\bar \s^{\n]\r}\bar\psi_\n\nonumber\\
    & +r_0\,\partial_\mu[\bar\psi_\nu\bar\s^\mu\s^\nu(\bar{\mathcal{G}}+\bar\r_1)]-ir_2\psi_\m\s^\m(\bar{\mathcal{G}}+\bar\r_1) -2\,{\bf{\L}}_2(\mathcal{G}-\r_1)\big[\tfrac{i}{2}\,\boldsymbol\o_\m h^\m{}_\r\s^\r(\bar{\mathcal{G}}+\bar\r_1)\nonumber\\
    & -i\o_\m^{(1)}\s^\m(\bar{\mathcal{G}}+\bar\r_1)\big] -\tfrac12 r_2\,\partial_\m(t^\m\partial_\n t^\n) -\tfrac16{\bf{\L}}_2 b_\m \s^\m(\mathcal{G}\bar{\mathcal{G}}+\mathcal{G}\bar\r_1-3\r_1\bar{\mathcal{G}}-3\r_1\bar\r_1) \nonumber\\
    &+\tfrac23\bold{\L}_2 M(\bar{\mathcal{G}}+\bar\r_1)^2+i\bold{\L}_2(\mathcal{G}-\r_1)\big[\pa_\m\big( t^\mu\boldsymbol\omega_\rho\sigma^\rho(\bar{\mathcal{G}}+\bar\r_1)\big)+\partial_\mu t^\mu\boldsymbol\omega_\rho\sigma^\rho(\bar{\mathcal{G}}+\bar\r_1)\big] \nonumber\\
    &-i\bold{\L}_2(\mathcal{G}-\r_1) t^\n\partial_\n(\boldsymbol{\o}_\m\s^\m)(\bar{\mathcal{G}}+\bar\r_1) -r_0\partial_\m[ (\mathcal{D}_\nu(\bar{\mathcal{G}}+\bar\rho_1))\bar{\sigma}^\mu\sigma^\nu(\bar{\mathcal{G}}+\bar\rho_1) \nonumber\\
    & -(\bar{\mathcal{G}}+\bar\rho_1)\bar\sigma^\mu \boldsymbol\omega_\nu\sigma^\nu(\bar{\mathcal{G}}+\bar\rho_1)]+h.c. \big\}\,.
\end{align}

\section{Goldstino Dynamics}\label{Goldstino Dynamics}
In order to zoom in on the goldstino dynamics in action \eqref{cubicaction}, we integrate out the auxiliary fields $b_\m, M$, $M^*$ and the Lagrange multipliers $r_0, r_0^*, \r_1, \bar\r_1, r_2$ and $r_2^*$. To simplify the process of integrating out the fields, we first perform suitable field redefinitions. \\ \\
We start by redefining the field $b_\mu$ as follows:
\be 
b_\mu=\mathcal{B}_\mu-\tfrac{1}{4}\mathcal{B}_\mu h+\tfrac{1}{4}\mathbf{\L}_2\mathcal{G}\sigma_\mu\bar{\mathcal{G}}-\tfrac{1}{4}\mathbf{\L}_2\r_1 \sigma_\mu \bar{\mathcal{G}}-\tfrac{1}{4}\mathbf{\L}_2\mathcal{G}\sigma_\mu\bar{\r}_1-\tfrac{3}{4}\mathbf{\L}_2\r_1\sigma_\mu \bar{\r}_1\,,
\ee 
such that the equation of motion of $\mathcal{B}_\mu$ is
\be
\mathcal{B}_\mu=0  +\mathcal{O}(3)\,. 
\ee 
Nex, we redefine $M$ as follows:
\be 
M=\mathcal{M}-\tfrac{1}{4}h\mathcal{M}+2\mathbf{\L}_2\mathcal{G}^2-2\mathbf{\L}_2 \r_1^2 +\tfrac{3}{2}t^\mu\partial_\mu+\tfrac{3}{2}r_0 \,,
\ee
and the equation of motion for $\mathcal{M}$ then becomes:
\be
\mathcal{M}=0 +\mathcal{O}(3)\,.
\ee 
Next redefine $r_2$ as:
\be 
r_2=\mathcal{R}_2+mr_0 \,.
\ee
Because of the boundary conditions \eqref{r0_bound_cond}, $\mathcal{R}_2$ vanishes at the boundary: 
\be\label{R2boundcond}
\mathcal{R}_2|_\infty=0\,.
\ee
Then the equation of motion of $r_0$ becomes
\be 
\label{eom_r0_after_r2_red}
\tfrac{3}{2}r_0^* +\tfrac{3}{2}r_0^*\left(\tfrac{1}{2}h-\partial_\mu t^\mu\right)+\hat{F}[r_0]=0\,,
\ee
where
\begin{align}
\hat{F}[r_0]=&\,\,\tfrac{2m^2}{3}(\bar{\mathcal{G}}+\bar{\r}_1)^2-\tfrac{2m^2}{3}(\mathcal{G}^2-\r_1^2)+2{\bf{\L}}_2(\bar{\mathcal{G}}^2-\bar{\r}^2_1) -im\psi_\mu\sigma^\mu(\bar{\mathcal{G}}+\bar{\r}_1 )\nonumber\\
    &
+\tfrac{4im}{3}\partial_\mu\mathcal{G}\sigma^\mu(\bar{\mathcal{G}}+\bar{\r}_1 )+\tfrac{2im}{3}\mathcal{G}\sigma^\mu\partial_\mu(2\bar{\mathcal{G}}-\bar{\r}_1 )-2im\r_1\sigma^\mu\partial_\mu\bar{\r}_1
+\bar{\psi}_\mu\bar{\sigma}^{\mu\nu}\bar{\psi}_\nu\nonumber\\
    &+\partial_\mu\!\left[\bar{\psi}_\nu\bar{\sigma}^\mu\sigma^\nu( \bar{\mathcal{G}}+\bar{\r}_1) \right]+\Box[( \bar{\mathcal{G}}+\bar{\r}_1)^2 ] +\partial_\m[2( \bar{\mathcal{G}}+\bar{\r}_1)\bar{\sigma}^\m\sigma^\n\partial_\n\bar{\r}_1\nonumber\\
    &-( \bar{\mathcal{G}}+\bar{\r}_1)\bar\s^\m(\pa_\n\s^\n)( \bar{\mathcal{G}}+\bar{\r}_1)-( \bar{\mathcal{G}}+\bar{\r}_1)\bar{\boldsymbol\o}_\n\bar\s^\m\s^\n( \bar{\mathcal{G}}+\bar{\r}_1)] \nonumber\\
&-\pa_\m\!\left[( \bar{\mathcal{G}}+\bar{\r}_1)\bar\s^\m\boldsymbol\o_\n\s^\n( \bar{\mathcal{G}}-\bar{\r}_1)\right]+\tfrac43 im\mathcal{G}\boldsymbol\o_\m\s^\m(\bar{\mathcal{G}}+\bar\r_1)-\tfrac23 im\mathcal{G}\s^\m\bar{\boldsymbol\o}_\m(2\bar{\mathcal{G}}-\bar\r_1)\,.
\end{align}
Now we redefine $r_0$ as
\be 
r_0=\mathcal{R}_0-\tfrac{1}{2}\left(\tfrac{1}{2}h-\partial_\mu t^\mu\right)\mathcal{R}_0 -\tfrac{2}{3}\hat{F_2}[r^*_0] \,,
\ee 
and the equation of motion for $\mathcal{R}_0$ is
\be
\mathcal{R}_0=0 +\mathcal{O}(3)\,,
\ee 
with the boundary condition
\be
\hat{F_2}[r_0] |_\infty =0\,.
\ee
Next, we find that the following field redefinitions serve to completely decouple $t^\mu$ from the fermions:
\be 
\begin{aligned}
\mathcal{G}=&\breve{\mathcal{G}}-\tfrac{1}{4}t^\mu\partial_\mu(\breve{\mathcal{G}}+\r_1)-\tfrac{1}{2\mathbf{\L}_2}\mathcal{R}_2(\breve{\mathcal{G}}+\rho_1)\,,\\
\r_1=&\mathcal{P}_1-t^\mu\partial_\mu \mathcal{P}_1\,.
\end{aligned}
\ee
From the boundary conditions \eqref{r0_bound_cond} we get
\be\label{P1_bound_cond}
\mathcal{P}_1|_\infty=0\,.
\ee 

The field $t^\m$ is redefined as follows:
\begin{align}
    \tilde t^\m=t^\m+\frac12 ht^\m-t^\m \partial_\n t^\n\,.
\end{align}
and the equation of motion for $\tilde t^\m$ is
\begin{align}
    \pa_\m Re[\mathcal{R}_2]=0\,.
\end{align}
Then using boundary condition \eqref{R2boundcond} we get 
\be\label{ReR2sol}
Re[\mathcal{R}_2]=0\,.
\ee
$\mathcal{R}_2$ and $\mathcal{R}^*_2$ have the same equation of motion which is
\be\label{eomR_2}
\begin{aligned}
&\partial_\mu \tilde t^\mu-\tfrac{1}{2}h -\tfrac{1}{4}\left(\tfrac{1}{2}h^2-h^{\mu\nu}h_{\mu\nu} \right)\\ &-\left[\tfrac{2m}{3}(\breve{\mathcal{G}}^2-\mathcal{P}_1^2)+i\partial_\mu(\breve{\mathcal{G}}-\mathcal{P}_1)\sigma^\mu(\bar{\breve{\mathcal{G}}} +\bar{\mathcal{P}}_1)+i(\breve{\mathcal{G}}-\mathcal{P}_1)\boldsymbol\o_\m\s^\m(\bar{\breve{\mathcal{G}}}+\bar{\mathcal{P}}_1) +h.c. \right]=0\,.
\end{aligned}
\ee
The e.o.m. for  $\mathcal{R}_2$ being the same as that of $\mathcal{R}^*_2$ implies that the imaginary part of $\mathcal{R}_2$ drops out of the action (which can also be checked explicitly). Now we look at the e.o.m. of $\mathcal{P}_1$. It is
\begin{align}
    4\mathbf{\L}_2\left(i\sigma^\mu{\bf D}_\mu\bar{\mathcal{P}}_1 -\tfrac{2}{3}m\mathcal{P}_1\right)+ \check{F}_1[\mathcal{P}_1,h_{\mu\nu}]+\check{F}_2[\bar{\mathcal{P}}_1,h_{\mu\nu}]=0\,,
\end{align}
where
\begin{align}
    &\check{F}_1[\mathcal{P}_1,h_{\mu\nu}]=-\,\frac43 m\boldsymbol{\L}_2h\mathcal{P}_1 \nonumber\\
    &\check{F}_2[\bar{\mathcal{P}}_1,h_{\mu\nu}]=2i\boldsymbol{\L}_2h\s^\m{ \bf D}_\m\bar{\mathcal{P}}_1-2i\boldsymbol{\L}_2h_{\m\n}\s^\m{\bf D}^\n\bar{\mathcal{P}}_1-4i\boldsymbol{\L}_2\s^\m\bar\o^{(1)}_\m\bar{\mathcal{P}}_1
\end{align}
Then we see that because of the boundary condition \eqref{P1_bound_cond}, we get
\be\label{R1sol}
\mathcal{P}_1=0 +\mathcal{O}(3)\,.
\ee
Now we plug the solutions for the auxiliary fields $\mathcal{B}_\m, \mathcal{M}$ and $\mathcal{M}^*$, and the solutions for the Lagrange multipliers $\mathcal{R}_0$, $\mathcal{R}^*_0$, $\mathcal{P}_1$, $\bar{\mathcal{P}}_1$ and $Re[\mathcal{R}_2]$, into the action \eqref{cubicaction}. Upon doing so the diffeomorphism St\"uckelberg field $t^\m$ automatically drops out of the action. We finally arrive at  the following action: 
\begin{align}\label{goldstino_action}
    S=\frac{1}{2\kappa^2}\int d^4x\,\big[\big\{&\sqrt{-g}\,R-\varepsilon^{\mu\nu\rho\lambda}\left(\psi_\m\sigma_\n \mathcal{D}_\r\bar\psi_\l+h.c.\right)\big\}^{(3)}-\big\{2\sqrt{-g}\,\left(\mathbf{\Lambda}_2-\frac{m^2}{3}\right)\big\}^{(3)} \nonumber \\
    &+\big\{\tfrac{2}{3} m\psi_\mu\sigma^{\mu\nu}\psi_\nu+2i\mathbf{\Lambda_2}\breve{\mathcal{G}}\sigma^\mu\bar{\psi}_\mu-2i\mathbf{\Lambda_2}\breve{\mathcal{G}}\sigma^\mu\mathbf{D}_\mu\bar{\breve{\mathcal{G}}}+\tfrac{4}{3} m\mathbf{\Lambda_2}\breve{\mathcal{G}}^2 \nonumber\\
    &+\,\tfrac13 hm\psi_\mu\sigma^{\mu\nu}\psi_\nu+\tfrac23m\psi_\mu h_\rho{}^{[\mu}{} \sigma^{\nu]\rho}\psi_\nu+\mathbf{\L}_2
    ( ih\breve{\mathcal{G}}\sigma^\mu\bar\psi_\mu -  i\breve{\mathcal{G}}h^\mu{}_\nu\sigma^\nu\bar \psi_\mu \nonumber \\
    & -ih\breve{\mathcal{G}}\sigma^\mu{\bf D}_\mu \bar{\breve{\mathcal{G}}}+i\breve{\mathcal{G}}h^\mu{}_\nu\sigma^\nu{\bf D}_\mu\bar{\breve{\mathcal{G}}}+2i\breve{\mathcal{G}}\sigma^\mu\bar\omega_\mu^{(1)}\bar{\breve{\mathcal{G}}}+\tfrac23hm\breve{\mathcal{G}}^2)+h.c.\big\}\big]\,.
\end{align}
This is the same as action \eqref{bandos3} which was obtained from \cite{Bandos:2015xnf}. This shows that up to the cubic order in fluctuations the minimal supergravity actions derived from the unimodular and nilpotent superfield approaches, are the same.

Action \eqref{goldstino_action} has terms only up to the leading order derivatives of the goldstino. We note that other combinations of terms realising supersymmetry nonlinearly can in principle be added to this supergravity action, while keeping it minimal (i.e.  without coupling to matter fields). However, they are expected to contain higher order derivatives of the goldstino \cite{Melville:2019tdc}, and may introduce ghost degrees of freedom.

\section{Discussion} \label{Disc}
 In this paper, we have initiated a comparison between two very different approaches to de Sitter supergravity. On the one hand, we have the earlier construction using nilpotent superfields \cite{Bergshoeff:2015tra}, and their elegant realisation as a spacetime filling $3$-brane coupled to the minimal $\mathcal{N}=1$ 4D supergravity\cite{Bandos:2015xnf}. On the other hand, we have the supersymmetric generalisation of unimodular gravity proposed in \cite{Nagy:2019ywi}. This includes a Lagrange multiplier chiral superfield that imposes a constraint on the chiral superspace.  Supersymmetry is spontaneously broken when the F-term in the Lagrange multiplier picks up a vev, allowing us to source a positive cosmological constant. At first glance, the two formulations are very different. For example, in the former the goldstino of the low energy effective theory is added explicitly. In the latter, it emerges implicitly through the excitations of a Stuckelberg field when the symmetry is spontaneously broken. But then as we showed in \autoref{Goldstino Dynamics}, at least up to the cubic order in fluctuations the two actions are related via field redefinitions.

To better understand the similarities and differences between the two models in \cite{Nagy:2019ywi} and \cite{Bandos:2015xnf}, it is instructive to look at the superfield version of the action for unimodular supergravity\cite{Nagy:2019ywi}:     
\be 
\label{full_action_after_stueck_disc}
\!\!\!S=-\tfrac{6}{8\pi G_N}\int d^4x\, d^2\Theta\big[\mathcal{E} \mathcal{R} + \tfrac{1}{6}\Lambda\big(\!\mathcal{E}-
\text{Ber}\!\left(\frac{\partial \Phi}{\partial X}\right) \mathcal{E}_0\left(\Phi\right)\!\big)  \!\big]        +h.c.\ ,
\ee
where
\be
\Lambda=\Lambda_0+\sqrt{2}\,\Theta \Lambda_1+\Lambda_2\Theta^2 \,.
\ee
The background solutions computed in \eqref{background_sol} suggest that we can define
\be 
\hat{\Lambda}=\Lambda_0-\tfrac{2m}{3}+\sqrt{2}\,\Theta \Lambda_1+\Lambda_2\Theta^2 \,,
\ee
such that
\be 
\Lambda=\hat{\Lambda}+\tfrac{2m}{3}\,.
\ee
Up to a constant term, the action can now be written as
\be 
\begin{aligned}
\!\!\!S=-\tfrac{12}{2 \kappa ^2}\int d^4x\,d^2\Theta
&\big[\mathcal{E} \big(\mathcal{R}+\tfrac{m}{9}\big)  
+ \tfrac{1}{6}\hat{\Lambda}\big(\!\mathcal{E}-
\text{Ber}\!\left(\frac{\partial \Phi}{\partial X}\right) \mathcal{E}_0\left(\Phi\right)\!\big)  \!\big]        +h.c.\ ,
\end{aligned}
\ee 
  This should be compared with  \eqref{bandosch} upon use of the relations in \autoref{App1}.  In each case, the first terms describe pure AdS supergravity, whereas the second gives the positive contributions to the cosmological constant. The supersymmetry breaking terms look very different, although we have now seen how they are equivalent up to cubic order in perturbations about the de Sitter vacuum solution, as shown in \autoref{unimod_sugra_cubic} and \autoref{Goldstino Dynamics}.
 
Given the  manifestly distinct starting points for the different constructions, this is an intriguing result. Moreover, in general, a locally supersymmetric action giving rise to dS solutions is not expected to be unique. An example of a term not present in our construction is:
\be\label{VSterm}
\sqrt{\tilde{e}}\,\tilde{g}^{\mu\nu}\tilde{\psi}_\mu\tilde{\psi}_\nu\,,
\ee
where we used the invariant vielbein and gravitino combinations 
\be 
\begin{aligned}
\tilde{e}_\mu^a&=e_\mu^a+\tilde{\mathcal{D}}_\mu X^a +2i\theta\sigma^a\bar{\psi}_\mu-2i\psi_\mu\sigma^a\bar{\theta}+i\theta\sigma^a\tilde{\mathcal{D}}_\mu\bar{\theta}-i\tilde{\mathcal{D}}_\mu\theta\sigma^a\bar{\theta}\,,\\
\tilde{\psi}_\mu&=\psi_\mu+\tilde{\mathcal{D}}_\mu\theta\,.
\end{aligned}
\ee 
These are constructed by Volkov-Soroka as in \cite{Volkov:1973jd,Volkov:1974ai,Volkov:1994te}, 
where $\tilde{\mathcal{D}}$ denotes a covariant derivative w.r.t.~an independent connection $\tilde\omega_\mu(x)$ introduced in \cite{Volkov:1973jd,Volkov:1974ai,Volkov:1994te}.
Note that $X^a$ is the St\"uckelberg field for local Poincar\'e translations. However, term \eqref{VSterm} is expected to contain a ghost degree of freedom (as can be seen by looking at the linearised approximation in \cite{VanNieuwenhuizen:1981ae}). The fact that both models failed to produce any of these (potentially pathological) terms might suggest a common origin in terms of  a consistent underlying theory. 

The equivalence between the two theories remains to be shown to higher order. If equivalence  can be shown, there would be compelling evidence for a more comprehensive (and perhaps more natural) construction, either at higher energies or higher $\mathcal{N}$. These superficially distinct formulations would be understood as following from different procedures for going to lower energies or (spontaneously) breaking the SUSY, but ultimately leading to the same physics. This putative theory would then have the potential to give an explanation at a more fundamental level for the different constraints employed in the models we looked at in this paper. On the other hand, if higher order effects reveal a fundamental difference in the two set-ups, it would be interesting to explore the phenomenological consequences in greater detail. We note with interest the recent application of nilpotent superfields as a natural way to project out certain Standard Model contributions to the vacuum energy \cite{tye}.  Could these methods also be adapted to a super-unimodular framework?

\section*{Acknowledgements}
We are grateful to Luca Martucci, Dmitri Sorokin and  David Stefanyszyn  for useful discussions. SB is supported by CUniverse research promotion project of Chulalongkorn University (grant CUAASC).  SN is supported by STFC grants ST/P000703/1 and ST/T000686/1. AP and IZ are both partially supported by STFC, through grants ST/S000437/1 and ST/P00055X/1 respectively.

\begin{appendix}

\vspace{.5in}

\section{Dictionary Between  \cite{Nagy:2019ywi} and  \cite{Bandos:2015xnf} }\label{App1}

Here we give  the dictionary between our notation which follows that of \cite{Nagy:2019ywi}, and the notation used in \cite{Bandos:2015xnf}, which we denote with a superscript $^{(B)}$. Note that we have different conventions:
\begin{align}
&R=-\,R^{(B)}\,, \nonumber \\
&\eta_{\m\n}{}=-\,\eta_{\m\n}{}^{(B)}\,,
\end{align}
and we use the map  
\begin{align}\label{dictionary}
&b_\mu{}=\frac38\,b_\mu{}^{(B)}\,, \nonumber \\
&M{}=-\frac34\,M{}^{(B)}\,, \nonumber \\
&\s_\m{}=-\,\s_\m{}^{(B)} \,,\nonumber \\
&\s_{\m\n}{}=\frac12\, \s_{\m\n}{}^{(B)}\,, \nonumber \\
&\psi_\m{}=-\,2\,\psi_\m{}^{(B)}\,, \nonumber \\
&\varepsilon=\varepsilon^{(B)} \quad \text{(where}\,\,\varepsilon\,\,\text{can have Lorentz, tangent space or spinor indices)}\,,\nonumber \\
&m=3\,m^{(B)} \,,\nonumber \\
&\mathbf{\L}_2 =\k^2f^2 \,,\nonumber \\
&\breve{\mathcal{G}}=\theta\,.
\end{align}

\section{Notations and Conventions}\label{App2}

Here we show our  conventions, which follow those in \cite{Wess:1992cp}. We work with the mostly plus signature metric $\eta_{\mu\nu}=\text{diag}(-1,1,1,1)$. $\a,\b,\g,\z,\dot\a,\dot\b,\dot\g,\dot\z$ are spinor indices taking values $1$ or $2$. All other Greek indices are Lorentz indices. 

\section*{Charge Conjugation Matrices}
\be
\varepsilon^{\a\b}=-\,\varepsilon_{\a\b}=\varepsilon^{\dot\a\dot\b}=-\,\varepsilon^{\dot\a\dot\b}=\bigl(\begin{smallmatrix}
0 &1 \\ 
-1 &0 
\end{smallmatrix}\bigr)\,,
\ee
\be
\varepsilon^{\a\b}\varepsilon_{\b\g}=\d^\a_\g\,,\quad \varepsilon^{\dot\a\dot\b}\varepsilon_{\dot\b\dot\g}=\d^{\dot\a}_{\dot\g}\,.
\ee

\section*{Pauli Matrices}

\be
\label{pauli_matrices}
\sigma^0=
\bigl(\begin{smallmatrix}
-1 &0 \\ 
0 &-1 
\end{smallmatrix}\bigr) ,\quad
\sigma^1=
\bigl(\begin{smallmatrix}
0 & 1\\ 
1 & 0
\end{smallmatrix}\bigr),\quad
\sigma^2=
\bigl(\begin{smallmatrix}
0 & -i\\ 
i & 0
\end{smallmatrix}\bigr), \quad
\sigma^3=
\bigl(\begin{smallmatrix}
1 & 0\\ 
0 & -1
\end{smallmatrix}\bigr)\,,
\ee
and 
\be
\bar{\sigma}^0=\sigma^0,\quad \bar{\sigma}^1=-\,\sigma^{1,2,3}\,,
\ee
\be
\s^\m=\s^\m{}_{\a\dot\b}\,, \quad \bar\s^\m=\bar\s^{\m\dot\a\b}\,.
\ee

\section*{Spin Connection}

\be 
\omega_\mu\equiv\omega_{\mu\alpha}^{\ \ \ \beta},\quad
\bar{\omega}_\mu\equiv \bar{\omega}_{\mu\ \ \dot{\beta}}^{\ \dot{\alpha}}\,,
\ee 
\be
\omega_{\m\a}{}^\b=-\,\frac12\,\o_{\m\n\r}\,(\s^{\n\r})_\a{}^\b=-\,\frac14\,\o_{\m\n\r}\,\s_{\a\dot\a}{}^\n\,\bar\s^{\r\dot\a\b}\,.
\ee
We note that we have the following symmetry when both the spinor indices are at the same level (see (B.7) in \cite{Wess:1992cp}):
\be 
\omega_{\mu\alpha\beta}=\omega_{\mu\beta\alpha}\,, \quad \omega_{\mu\dot\alpha\dot\beta}=\omega_{\mu\dot\beta\dot\alpha}\,.
\ee

\section*{Raising and Lowering of Spinor Indices}

\be
\chi^\a=\varepsilon^{\a\b}\chi_\b\,,\quad\chi_\a=\varepsilon_{\a\b}\chi^\b\,,\quad\bar\chi_{\dot\a}=\varepsilon_{\dot\a \dot\b}\,\bar\chi^{\dot\b}\,,\quad\bar\chi^{\dot\a}=\varepsilon^{\dot\a \dot\b}\bar\chi_{\dot\b}\,,
\ee
\be
\bar\s^{\m\dot\a\b}=\varepsilon^{\dot\a\dot\z}\varepsilon^{\b\g}\s^\m{}_{\g\dot\z}\,,\quad\s^\m{}_{\a\dot\b}=\varepsilon_{\a\z}\,\varepsilon_{\dot\b\dot\g}\,\bar\s^{\m\dot\g\z}\,,
\ee
\be
\begin{aligned}
\o_\m{}^\a{}_\b&=\varepsilon^{\a\g}\,\o_{\m\g\b}=\varepsilon^{\a\g}\,\o_{\m\b\g}=\o_{\m\b}{}^\a \\
&=\varepsilon_{\b\g}\,\o_\m{}^{\a\g}=\varepsilon_{\b\g}\,\o_\m{}^{\g\a}=\o_{\m\b}{}^\a\,,
\end{aligned}
\ee
\be
\begin{aligned}
\bar\o_{\m\dot\a}{}^{\dot\b}&=\varepsilon^{\dot\b\dot\g}\,\bar\o_{\m\dot\a\dot\g}=\varepsilon^{\dot\b\dot\g}\,\bar\o_{\m\dot\g\dot\a}=\bar\o_\m{}^{\dot\b}{}_{\dot\a} \\
&=\varepsilon_{\dot\a\dot\g}\,\bar\o_\m{}^{\dot\g\dot\b}=\varepsilon_{\dot\a\dot\g}\,\bar\o_\m{}^{\dot\b\dot\g}=\bar\o_\m{}^{\dot\b}{}_{\dot\a}\,,
\end{aligned}
\ee
\underline{Some Identities}:
\be
\chi\s^\m\bar\psi=-\,\bar\psi\bar\s^\m\chi\,, \quad \chi\s^\m\bar\s^\n\psi=\psi\s^\n\bar\s^\m\chi\,.
\ee

\section*{Hermitian Conjugation}

\be
\chi=\begin{bmatrix}
        \chi_1 \\
        \chi_2
    \end{bmatrix}\,, \quad
\chi^*=\begin{bmatrix}
        \chi^*_1 \\
        \chi^*_2
        \end{bmatrix}\,, \quad
\chi^\dagger=(\chi^*)^T=\begin{bmatrix}
        \chi^*_1 & \chi^*_2
        \end{bmatrix}\,,
\ee
\be
(\chi^\a)^*=\bar\chi^{\dot\a}\,, \quad (\bar\chi^{\dot\a})^*=\chi^\a\,, \quad (\chi_\a)^*=\bar\chi_{\dot\a}\,, \quad (\bar\chi_{\dot\a})^*=\chi_\a\,,
\ee
\be
(\chi\psi)^\dagger=(\chi^\a\psi_\a)^\dagger=(\chi^\a\psi_\a)^*=\psi^*_\a\,\chi^{\a*}=\bar\psi_{\dot\a}\,\bar\chi^{\dot\a}=\bar\psi\bar\chi\,,
\ee
\be
(\s^\m{}_{\a\dot\b})^*=\s^\m{}_{\b\dot\a}\,,\quad (\bar\s^\m{}^{\dot\a\b})^*=\bar\s^{\m\dot\b\a}\,,
\ee
\be
(\omega_{\mu\alpha}^{\ \ \ \beta})^*=-\,\bar\o_\m{}^{\dot\b}{}_{\dot\a}\,,\quad (\bar\o_\m{}^{\dot\b}{}_{\dot\a})^*=-\,\omega_{\mu\alpha}^{\ \ \ \beta}\,.
\ee
\underline{Some Identities}:
\be
(\chi\s^\m\bar\psi)^\dagger=\psi\s^\m\bar\chi=-\,\bar\chi\bar\s^\m\psi\,, \quad (\chi\s^\m\bar\s^\n\psi)^\dagger=\bar\psi\bar\s^\n\s^\m\bar\chi=\bar\chi\bar\s^\m\s^\n\bar\psi\,,
\ee
\be
(\psi\,\o_\m\s^\m\,\bar\chi)^\dagger=-\,\chi\,\s^\m\bar\o_\m\,\bar\psi\,.
\ee

\section*{Covariant Derivative}
\be 
\begin{aligned}
\mathcal{D}_\mu\chi^\alpha&=\partial_\mu\chi^\alpha+\chi^\beta\omega_{\mu\beta}^{\ \ \ \alpha}\equiv \partial_\mu\chi+\chi\omega_\mu \,,\\
\mathcal{D}_\mu\bar{\chi}_{\dot{\alpha}}&=\partial_\mu\bar{\chi}_{\dot{\alpha}}+\bar{\chi}_{\dot{\beta}}\bar{\omega}_{\mu\ \ \dot{\alpha}}^{\ \dot{\beta}}\equiv \partial_\mu\bar{\chi} + \bar{\chi}\bar{\omega}_\mu \,.
\end{aligned}
\ee 
Following are the expressions for the opposite positions of the spinor index:
\be 
\begin{aligned} 
\mathcal{D}_\mu\chi_\alpha&=\partial_\mu\chi_\alpha+\chi^\beta\omega_{\mu\beta\alpha}=\partial_\mu\chi_\alpha+\omega_{\mu\alpha\beta}\chi^\beta=\partial_\mu\chi_\alpha-\omega_{\mu\alpha}^{\ \ \ \beta}\chi_\beta\equiv\partial_\mu\chi-\omega_\mu\chi\,,\\
\mathcal{D}_\mu\bar{\chi}^{\dot{\alpha}}&=\partial_\mu\bar{\chi}^{\dot{\alpha}}+\bar{\chi}_{\dot{\beta}}\bar{\omega}_\mu^{\ \dot{\beta}\dot{\alpha}}=\partial_\mu\bar{\chi}^{\dot{\alpha}}+\bar{\omega}_\mu^{\ \dot{\alpha}\dot{\beta}}\bar{\chi}_{\dot{\beta}}=\partial_\mu\bar{\chi}^{\dot{\alpha}}-\bar{\omega}_{\mu\ \ \dot{\beta}}^{\ \dot{\alpha}}\bar{\chi}^{\dot{\beta}}\equiv\partial_\mu\bar{\chi}-\bar{\omega}_\mu\bar{\chi}\,.
\end{aligned}
\ee 

\section*{Covariant Derivative of Pauli Matrix} 
The covariant derivative of the Pauli matrix vanishes. Using the identities above, we can write it as
\be 
\mathcal{D}_\mu \sigma^\nu= \nabla_\mu\sigma^\nu-\omega_\mu\sigma^\nu +\sigma^\nu\bar{\omega}_\mu=0\,,
\ee 
with $\nabla=\partial+\Gamma$. Taking the trace of the above, and using identity (7.113) from \cite{Freedman:2012zz}
\be
\sqrt{-g}\,\nabla_\mu V^\mu=\partial_\mu\left(\sqrt{-g}\, V^\mu\right)\,,
\ee 
together with the unimodularity condition, we have, at zeroth order in pertubation:
\be 
{\bf D}_\mu \sigma^\mu= \partial_\mu\sigma^\mu-\boldsymbol\omega_\mu\sigma^\mu +\sigma^\mu\bar{\boldsymbol\omega}_\mu=0\,,
\ee
and at first order:
\be  
\tfrac{1}{2}\left(\partial_\nu h -\partial_\mu h^\mu_{\nu} \right)\sigma^\nu-\tfrac{1}{2}h^\mu_{\rho}\left(\partial_\mu\sigma^\rho-\boldsymbol{\omega}_\mu\sigma^\rho+\sigma^\rho\bar{\boldsymbol\omega}_\mu \right)
-\omega^{(1)}_\mu\sigma^\mu+\sigma^\mu\bar{\omega}_\mu^{(1)}=0 \,.
\ee

\section*{Superfield ${\cal R}$}

The chiral supergravity superfield ${\mathcal R}$ in component form is:

\be\label{Rsuperfield}
\begin{aligned}
{\cal R}=&-\tfrac{1}{6}\{M+\Theta(\sigma^\mu\bar{\sigma}^\nu\psi_{\mu\nu}-i\sigma^\mu\bar{\psi}_\mu M+i\psi_\mu b^\mu)
+\Theta^2[\tfrac{1}{2}R+i\bar{\psi}^\mu\bar{\sigma}^\nu\psi_{\mu\nu}+\tfrac{2}{3}MM^*
+\tfrac{1}{3}b^\mu b_\mu\\
&-ie_a^\mu\,\mathcal{D}_\mu b^a+\tfrac{1}{2}\bar{\psi}\bar{\psi}M-\tfrac{1}{2}\psi_\mu\sigma^\mu\bar{\psi}_\nu b^\nu +\tfrac{1}{8}\varepsilon^{\mu\nu\rho\sigma}(\bar{\psi}_\mu\bar{\sigma}_\nu\psi_{\rho\sigma}+\psi_\mu\sigma_\nu\bar{\psi}_{\rho\sigma})
]\}\,.
\end{aligned} 
\ee

\section*{Secondary Fluctuations}
Following are the fluctuations used in this work that are derived from the primary fluctuations shown in \eqref{fluct}:
\be
\begin{aligned}
\sqrt{-g}&=1+\tfrac{1}{2}h+\tfrac{1}{4}\left(\tfrac{1}{2}h^2-h_{\mu\nu}h^{\mu\nu}\right)\,,\\
\s_\m&=\s_\m+\frac12\, h_\m{}^\r\s_\r \,,\\
\s^\m&=\s^\m-\frac12\, h^\m{}_\r\s^\r \,,\\
\s_{\m\n}&=\s_{\m\n}-h_{[\m}{}^\r\s_{\n]\r} \,,\\
\s^{\m\n}&=\s^{\m\n}+h^{[\m}{}_\r\s^{\n]\r} \,.
\end{aligned}
\ee

\end{appendix}

\addcontentsline{toc}{section}{References}
\bibliographystyle{utphys}

\bibliography{refs}

\end{document}